\documentclass[11pt]{article}%

\usepackage{tikz}
\usepackage{adjustbox}

\usepackage[latin1]{inputenc}
\usepackage{natbib}

\usepackage[english]{babel}
\usepackage{pst-all}
\usepackage{amssymb}

\usepackage{amsmath,enumerate,rotate}
\usepackage{amssymb,latexsym}
\usepackage{epsfig}
\usepackage{graphicx, graphics,float,subfigure}
\usepackage{amsthm}

\usepackage{multirow}
\usepackage{tabularx,ragged2e}
\usepackage{booktabs}

\usepackage[colorlinks=true,linkcolor=blue]{hyperref}%
\usepackage{graphicx}
\hypersetup{urlcolor=blue, citecolor=blue, colorlinks=true, linkcolor=blue} 
\usepackage[margin=1in]{geometry}
\usepackage{verbatim}

\floatstyle{plaintop}
\restylefloat{table}
\usepackage[tableposition=top]{caption}
 \usepackage{mathtools}
\DeclarePairedDelimiter\ceil{\lceil}{\rceil}

\usepackage{soul}
\usepackage{color}
\usepackage[mathcal]{eucal}
\usepackage{mathrsfs} 
\usepackage{verbatim}
\usepackage{url}
\usepackage{setspace}
\usepackage{mathtools} 

\newcommand\arc{\operatorname{{\rm{arc}}}}
\newcommand{\sinc}{\operatorname{\rm sinc}}

\theoremstyle{plain}
\newtheorem{theorem}{Theorem}
\newtheorem*{theorem*}{Theorem}
\newtheorem{lemma}[theorem]{Lemma}

\newtheorem*{corollary*}{Corollary}
\newtheorem*{definition*}{Definition}

\theoremstyle{remark}

\def\br{\boldsymbol{r}}
\def\bx{\boldsymbol{x}}
\def\by{\boldsymbol{y}}
\def\bs{\boldsymbol{s}}
\def\S{\mathbb{S}}
\def\R{\mathbb{R}}
\def\Z{\mathbb{Z}}
\def\I{\mathbb{I}}
\def\E{\mathbb{E}}
\def\X{\mathbb{X}}
\def\V{\mathbb{V}}
\def\bC{\boldsymbol{C}}
\def\bZ{\boldsymbol{Z}}
\def\bX{\boldsymbol{X}}
\def\bY{\boldsymbol{Y}}
\def\bs{\boldsymbol{s}}
\def\bh{\boldsymbol{h}}
\def\bm{\mathbf}

\newcommand{\blam}{ \mbox{\boldmath $ \lambda $} }
\newcommand{\bet}{ \mbox{\boldmath $ \eta $} }
\newcommand{\bome}{ \mbox{\boldmath $ \omega $} }
\newcommand{\bbet}{ \mbox{\boldmath $ \beta $} }
\newcommand{\bbeta}{ \mbox{\boldmath $ \beta $} }
\newcommand{\balph}{ \mbox{\boldmath $ \alpha $} }
\newcommand{\balpha}{ \mbox{\boldmath $ \alpha $} }
\newcommand{\bphi}{ \mbox{\boldmath $\phi$}}
\newcommand{\bzeta}{ \mbox{\boldmath $\zeta$}}
\newcommand{\bkap}{ \mbox{\boldmath $\kappa$}}
\newcommand{\bkappa}{ \mbox{\boldmath $\kappa$}}
\newcommand{\beps}{ \mbox{\boldmath $\epsilon$}}
\newcommand{\bepsilon}{ \mbox{\boldmath $\epsilon$}}
\newcommand{\bthet}{ \mbox{\boldmath $ \theta $} }
\newcommand{\btheta}{ \mbox{\boldmath $ \theta $} }
\newcommand{\bnu}{ \mbox{\boldmath $\nu$} }
\newcommand{\bmu}{ \mbox{\boldmath $\mu$} }
\newcommand{\bOmega}{ \mbox{\boldmath $\Omega$} }
\newcommand{\bGam}{ \mbox{\boldmath $\Gamma$} }
\newcommand{\bSig}{ \mbox{\boldmath $\Sigma$} }
\newcommand{\bSigma}{ \mbox{\boldmath $\Sigma$} }
\newcommand{\bPhi}{ \mbox{\boldmath $\Phi$} }
\newcommand{\bThet}{ \mbox{\boldmath $\Theta$} }
\newcommand{\bTheta}{ \mbox{\boldmath $\Theta$} }
\newcommand{\bDel}{ \mbox{\boldmath $\Delta$} }
\newcommand{\bDelta}{ \mbox{\boldmath $\Delta$} }
\newcommand{\bnabla}{ \mbox{\boldmath $\nabla$} }
\newcommand{\bLam}{ \mbox{\boldmath $\Lambda$} }
\newcommand{\bLambda}{ \mbox{\boldmath $\Lambda$} }
\newcommand{\bLambdasub}{ \scriptsize{\bLambda}}
\newcommand{\bgam}{ \mbox{\boldmath $\gamma$} }
\newcommand{\bgamma}{ \mbox{\boldmath $\gamma$} }
\newcommand{\brho}{ \mbox{\boldmath $\rho$} }
\newcommand{\bdel}{ \mbox{\boldmath $\delta$} }
\newcommand{\bdelta}{ \mbox{\boldmath $\delta$} }
\newcommand{\bvarphi}{ \mbox{\boldmath $\varphi$} }
\newcommand{\bsigma}{ \mbox{\boldmath $\sigma$} }
\newcommand{\boeta}{ \mbox{\boldmath $\eta$} }

\newcommand{\betSS}{\scriptsize\bet}
\newcommand{\bepsilonSS}{\scriptsize\bepsilon}

\newcommand{\bzero}{\textbf{0}}
\newcommand{\bone}{\textbf{1}}

\newcommand{\bz}{\textbf{z}}
\newcommand{\ba}{\textbf{a}}
\newcommand{\bA}{\textbf{A}}
\newcommand{\bb}{\textbf{b}}
\newcommand{\bc}{\textbf{c}}
\newcommand{\bd}{\textbf{d}}
\newcommand{\bbf}{\textbf{f}}
\newcommand{\bk}{\textbf{k}}
\newcommand{\bK}{\textbf{K}}
\newcommand{\bH}{\textbf{H}}
\newcommand{\bi}{\textbf{i}}
\newcommand{\bI}{\textbf{I}}
\newcommand{\bg}{\textbf{g}}
\newcommand{\bG}{\textbf{G}}
\newcommand{\bJ}{\textbf{J}}
\newcommand{\bL}{\textbf{L}}
\newcommand{\bM}{\textbf{M}}
\newcommand{\bn}{\textbf{N}}
\newcommand{\bO}{\textbf{O}}
\newcommand{\bp}{\textbf{p}}
\newcommand{\bt}{\textbf{t}}
\newcommand{\bu}{\textbf{u}}
\newcommand{\bv}{\textbf{v}}
\newcommand{\bV}{\textbf{V}}
\newcommand{\bw}{\textbf{w}}

\renewcommand\Re{\operatorname{\rm{Re}}}
\renewcommand\Im{\operatorname{\rm{Im}}}

\DeclareMathOperator{\esssup}{\mathrm{ess\: sup}}
\DeclareMathOperator{\essinf}{\mathrm{ess\: inf}}
\DeclareMathOperator{\sgn}{sgn}
\DeclareMathOperator{\mean}{\mathbb{E}}
\DeclareMathOperator{\cov}{Cov}
\DeclareMathOperator{\var}{Var}

\newcommand\norm[1]{{\left\Vert{#1}\right\Vert}}
\newcommand\itemno[1]{(\romannumeral #1)}
\newcommand\set[1]{{\left\{#1\right\}}}
\newcommand\sups[1]{\sup{\left\{#1\right\}}} 
\newcommand\infset[1]{\inf{\left\{#1\right\}}}
\newcommand\bigset[1]{\bigl\{#1\bigr\}}
\newcommand\Bigset[1]{\Bigl\{#1\Bigr\}}
\newcommand\vet[1]{{\mathbf{#1}}}
\newcommand\bigmod[1]{\bigl\vert{#1}\bigr|}
\newcommand\Bigmod[1]{\Bigl\vert{#1}\Bigr|}
\newcommand\biggmod[1]{\biggl\vert{#1}\biggr|}
\newcommand\bignorm[2]{\left.{\bigl\Vert{#1}\bigr\Vert_{#2}}\right.}
\newcommand\Bignorm[2]{\left.{\Bigl\Vert{#1}\Bigr\Vert_{#2}}\right.}
\newcommand\biggnorm[2]{\left.{\biggl\Vert{#1}\biggr\Vert_{#2}}\right.}
\newcommand\opnorm[2]{|\!|\!| {#1} |\!|\!|_{#2}}
\newcommand\Bigopnorm[2]{\left.{\Big|\!\Big|\!\Big| {#1} \Big|\!\Big|\!\Big|_{#2}}\right.}
\newcommand\abs[1]{\left\vert{#1}\right\vert}
\newcommand\ds{\displaystyle}

\newcommand\tonde[1]{\left(#1\right)}

\newcommand\Smallfrac[2]{\mbox{\footnotesize$\displaystyle\frac{#1}{#2}$}}
\newcommand\smallfrac[2]{\mbox{\small$\displaystyle\frac{#1}{#2}$}}

\newcommand\virgolette[1]{`#1'}
\newcommand\virgolettedoppie[1]{``#1''}
\newcommand\tsum{\textstyle\sum}
\newcommand\tprod{\textstyle\prod}
\newcommand\diff{\mathrm{d}}
\newcommand\duerighe[2]{\genfrac{}{}{0pt}{}{#1}{#2}}
\newcommand\ind{\mathbb{I}}
\newcommand\pspace{\mathbb{P}}
\newcommand\psigmaalg{\mathscr{P}}
\newcommand\mle{{\hat \te}_n}
\newcommand\prob{\mathbb{P}}
\newcommand\Mean[1]{\mathbb{E}\left(#1\right)}

\newcommand\rfact[2]{\left(#1\right)^{(#2)}}

\newcommand\To{\longrightarrow}
\newcommand\Borel{\mathscr{B}}

\newcommand\bB{\mathbf{B}}
\newcommand\bD{\mathbf{D}}
\newcommand\bE{\mathbf{E}}
\newcommand\bF{\mathbf{F}}
\newcommand\bN{\mathbb{N}}
\newcommand\bR{\mathbb{R}}
\newcommand\bP{\mathbf{P}}
\newcommand\bS{\mathbf{S}}
\newcommand\bT{\mathbf{T}}
\newcommand\bU{\mathbf{U}}
\newcommand\bW{\mathbf{W}}

\newcommand\BB{\mathbb{B}}
\newcommand\BC{\mathbb{C}}
\newcommand\BD{\mathbb{D}}
\newcommand\BE{\mathbb{E}}
\newcommand\BF{\mathbb{F}}
\newcommand\BN{\mathbb{N}}
\newcommand\BR{\mathbb{R}} \newcommand\BRd{\mathbb{R}^d}
\newcommand\BP{\mathbb{P}}
\newcommand\BQ{\mathbb{Q}}
\newcommand\BS{\mathbb{S}}
\newcommand\BT{\mathbb{T}}
\newcommand\BU{\mathbb{U}}
\newcommand\BW{\mathbf{W}}
\newcommand\BZ{\mathbb{Z}}

\newcommand\cA{\mathcal{A}}  \newcommand\fA{\mathfrak{A}}
\newcommand\cB{\mathcal{B}}  \newcommand\fB{\mathfrak{B}}
\newcommand\cC{\mathcal{C}}  \newcommand\fC{\mathfrak{C}}
\newcommand\cD{\mathcal{D}}
\newcommand\cE{\mathcal{E}}  \newcommand\fE{\mathfrak{E}}
\newcommand\cF{\mathcal{F}}  \newcommand\fF{\mathfrak{F}}
\newcommand\cG{\mathcal{G}}
\newcommand\cH{\mathcal{H}}  \newcommand\fH{\mathfrak{H}}
\newcommand\cI{\mathcal{I}}  \newcommand\fI{\mathfrak{I}}
\newcommand\cJ{\mathcal{J}}
\newcommand\cK{\mathcal{K}}  \newcommand\fK{\mathfrak{K}}
\newcommand\cL{\mathcal{L}}  \newcommand\fL{\mathfrak{L}}
\newcommand\cM{\mathcal{M}}  \newcommand\fM{\mathfrak{M}}
\newcommand\cN{\mathcal{N}}  \newcommand\fN{\mathfrak{N}}
\newcommand\cP{\mathcal{P}}  \newcommand\fP{\mathfrak{P}}
\newcommand\cQ{\mathcal{Q}}
\newcommand\cR{\mathcal{R}}  \newcommand\fR{\mathfrak{R}}
\newcommand\cS{\mathcal{S}}  \newcommand\fS{\mathfrak{S}}
\newcommand\cT{\mathcal{T}}  \newcommand\fT{\mathfrak{T}}
\newcommand\cU{\mathcal{U}}  \newcommand\fU{\mathfrak{U}}
\newcommand\cW{\mathcal{W}}  \newcommand\fW{\mathfrak{W}}
\newcommand\cX{\mathcal{X}}  \newcommand\fX{\mathfrak{X}}
\newcommand\cY{\mathcal{Y}}

\newcommand\wB{{\widetilde B}}
\newcommand\wq{{\widetilde q}}

\newcommand\al{\alpha}
\newcommand\be{\beta}
\newcommand\ga{\gamma}    \newcommand\Ga{\Gamma}
\newcommand\de{\delta}
\newcommand\ep{\epsilon}  \newcommand\vep{\varepsilon}
\newcommand\ka{{\kappa}}
\newcommand\la{\lambda}   \newcommand\La{\Lambda}
\newcommand\om{\omega}    \newcommand\Om{\Omega}  \newcommand\Omi{{\rm O}}
\newcommand\te{\theta}
\newcommand\Te{\Theta}
\newcommand\vp{\varphi}
\newcommand\ze{\zeta}
\newcommand\varte{\vartheta}

\newcommand\ta{{\tilde a}}
\newcommand\tc{{\tilde c}}
\newcommand\tK{{\tilde K}}
\newcommand\tN{{\tilde N}}

\def\simind{\stackrel{\mbox{\scriptsize{ind}}}{\sim}}
\def\simiid{\stackrel{\mbox{\scriptsize{iid}}}{\sim}}

\newcommand\ppy{\al}

\newcommand\prior{\mathbb{P}}
\newcommand\post[1]{\mathbb{P}^{#1}}

\newcommand\legendre{\bar{P}^m_{n}(\sin L)}
\newcommand\legendree{\bar{P}^m_{n'}(\sin L')}

\allowdisplaybreaks
\begin{document}

	

\fontsize{12}{14pt plus.8pt minus .6pt}\selectfont \vspace{0.8pc}

\vspace{2pt} \centerline{\LARGE \bf Nonseparable Space-Time Stationary Covariance}
\centerline{\LARGE \bf Functions on Networks cross Time}
\vspace{.3cm} 
\begin{center}
	{\large {\sc Emilio Porcu}}\footnote{\baselineskip=10pt
	    Department of Mathematics, 
	    Khalifa University, Abu Dhabi, United Arab Emirates, 
		$\&$ School of Computer Science and Statistics, 
		Trinity College Dublin, Dublin, Ireland
 \\
		E-mail: emilio.porcu@ku.ac.ae  
	}, 
{\large {\sc Philip White}}\footnote{\baselineskip=10pt
		Department of Statistics, Brigham Young University, Provo, Utah, USA \\
		E-mail: pwhite@stat.byu.edu
	} and {\large {\sc Marc G. Genton}}\footnote{\baselineskip=10pt
		Statistics Program, King Abdullah University of Science and Technology, Thuwal 23955-6900, Saudi Arabia E-mai: marc.genton@kaust.edu.sa
	} 
	
\vspace{1.3cm} 
\today
\vspace{1.3cm} 

\begin{abstract}
The advent of data science has provided an increasing number of challenges with high data complexity.
This paper addresses the challenge of space-time data where the spatial domain is not a planar surface, a sphere, or a linear network, but a generalized network (termed a graph with Euclidean edges). Additionally, data are repeatedly measured over different temporal instants. 
We provide new classes of nonseparable space-time stationary covariance functions where {\em space} can be a generalized network, a Euclidean tree, or a linear network, and where time can be linear or circular (seasonal). 
Because the construction principles are technical, we focus on illustrations that guide the reader through the construction of statistically interpretable examples. A simulation study demonstrates that we can recover the correct model when compared to misspecified models. In addition, our simulation studies show that we effectively recover simulation parameters. In our data analysis, we consider a traffic accident dataset that shows improved model performance based on covariance specifications and network-based metrics.

\vspace{4cm}

\end{abstract}
\end{center}


\newpage

\baselineskip 25 pt

\section{Introduction}

\subsection{Context and State of the Art}

The data science revolution has introduced many challenges intimately related with data complexity. Amongst many, challenges related to domain complexity in geo-referenced data or point processes cover an important part of the literature, and the reader is referred to \cite{anderes2020}, \cite{moradi}, \cite{BADDELEY2021100435} and \cite{BADDELEY} for recent contributions. 

This paper deals with covariance functions for space-time Gaussian random fields, with space either a generalized (or a linear) network, or a Euclidean tree, and time either linear (the real line) or circular (the unit circle). The apparent  importance of stochastic processes defined over generalized networks is shown by an increasing number of applications in spatial statistics \citep{cressie2006spatial, gardner2003predicting, ver2006spatial, peterson2013modelling, peterson2007geostatistical, montembeault2012impact}, point processes \citep{xiao2017modeling, perry2013point, deng2014ginibre, baddeley2017stationary}, and machine learning \citep{alsheikh2014machine, georgopoulos2014distributed,hamilton2017representation}. 

For Gaussian random fields and Gaussian process modeling, inference and prediction, the covariance function plays a crucial role. Covariance functions are positive definite, and such a requirement is non trivial to check. Building positive definite functions is even more challenging for stochastic processes defined over networks, and we refer the reader to \cite{ver2006spatial} and \cite{peterson2007geostatistical} for efforts in this direction. 

Spectral techniques are often used to check for positive definiteness of a candidate function. Unfortunately, no spectral representations are available for the cases studied in this paper, making the problem even more challenging. Actually, even the definition of stationarity over networks is controversial \citep{baddeley2017stationary}. These facts motivated  \cite{anderes2020} to consider covariance functions being isotropic over generalized networks. Hence, the covariance function depends on the distance between any two points located over the network. 

To generalize a linear network, \cite{anderes2020} proposed graphs with Euclidean edges: graphs where each edge is associated to an abstract set that is in bijective correspondence with a segment of the real line. This provides each edge with a Cartesian coordinate system to measure distances between any two points on that edge.

\subsection{The Problems and our Contributions}

\cite{anderes2020} constructed generalized linear networks using two alternative metrics for graphs with Euclidean edges. Further, they provided sufficient conditions for function classes to be positive definite when composed with such metrics. 

Simple generalizations to the setting where generalized networks are considered as topological structures that do not evolve over time are provided by \cite{tang}. In particular, \cite{tang} adapted a version of the Gneiting class \citep{Gneiting:2002} to generalized networks cross linear time. Unfortunately, in their formulation the temporal distance is rescaled by spatial component, while Gneiting's original class proposed the opposite (rescaling spatial distance with temporal variables), which is often more realistic for geostatistical applications. Therefore, further investigations are needed.

The literature on space-time covariance functions with space being a network is minimal. There is a clear lack of options for covariance functions that describe a wealth of interactions between space and time. This paper provides a very general class of nonseparable space-time covariance functions where space can be a generalized or a linear network, or a Euclidean tree. As for time, special emphasis will be put on time considered as linear, albeit our contribution extends, under some mild regularity conditions, to circular (seasonal) time as well.

The new class of covariance functions allows for either the geodesic or resistance metrics in space (details given subsequently). Further, the proposed structure is general enough to describe several types of interactions between space and time. We also allow for compactly supported covariance functions when working on Euclidean trees. 

Specifically, we provide the following contributions:
\begin{itemize}
\item[a)] We describe a procedure to compose two parametric classes of functions describing, respectively, spatial and temporal dependence. Such a combination provides the new class of nonseparable space-time covariance functions.
\item[b)] A special case of this class results in an adaptation of the Gneiting class of covariance functions, that had been originally proposed over planar surfaces \citep{Gneiting:2002,Porcu20111293}. 
\item[c)] We consider covariance functions that are dynamically compactly supported over generalized networks cross linear or circular time. This means that for every fixed temporal lag the covariance functions are compactly supported over balls with given radii. This allows important computational gains. 
\item[d)] A simulation study addresses three aspects. We work on a generalized network and compare correctly specified and misspecified models. First, we show the impact of using the incorrect distance metric in terms of likelihood estimation. Second, identifiabilty problems are inspected. Third, we verify that model performance is best under a correct choice of the spatial metric and correct covariance function. 
\item[e)] We analyze a traffic accident point pattern dataset. Comparing various models that differ in terms of covariance function, distance metric, and probability mass function. We find that the best model uses the network distance and a covariance model developed in our paper. Using this model, we explore the estimated space-time random effect and associated correlation structure.
\end{itemize}

The results obtained in this paper are technical and require substantial background on quasi metric spaces, isometric embeddings, graphs with Euclidean edges, and harmonic analysis. Hence, the exposition focuses on describing the main ideas and illustrating their implications through practical examples, while keeping a statistical language and deferring the technical part to the Supplemental Material. The plan of the paper is the following. Section \ref{sec2} contains a succinct statistical background and an illustration about graphs with Euclidean edges. Section \ref{sec3a} describes the main idea and construction for the new class of space-time covariance functions. Section \ref{sec3b} guides the reader through practical examples. We also discuss previously proposed examples in concert with models that are not valid. A simulation study in Section \ref{sec4} illustrates the practical implementation of our model along  with the statistical implications of working with the correct set of spatial distances. Section \ref{sec5} analyzes a traffic accident point pattern dataset and compares various models that differ in terms of covariance function, distance metric, and probability mass function. A short discussion concludes the paper. 

The Supplemental Material is technical and contains the following. All theorems referenced are stated and proved in the Supplemental Material. All Supplemental Material sections have an ``S'' prefix, while tables, figures, and equations are given number labels as if they followed subsequently in the manuscript. Section \ref{A} presents a mathematical background needed to understand the proofs. Section \ref{sec3} provides formal statements and their proofs to justify the general ideas illustrated in Section \ref{sec3a}. Section \ref{A.4} reports tables that allow to construct a wealth of practical examples of new covariance functions.


\section{Background and Notation} \label{sec2}

\begin{figure}[h!]
    \centering    \includegraphics[scale=0.55]{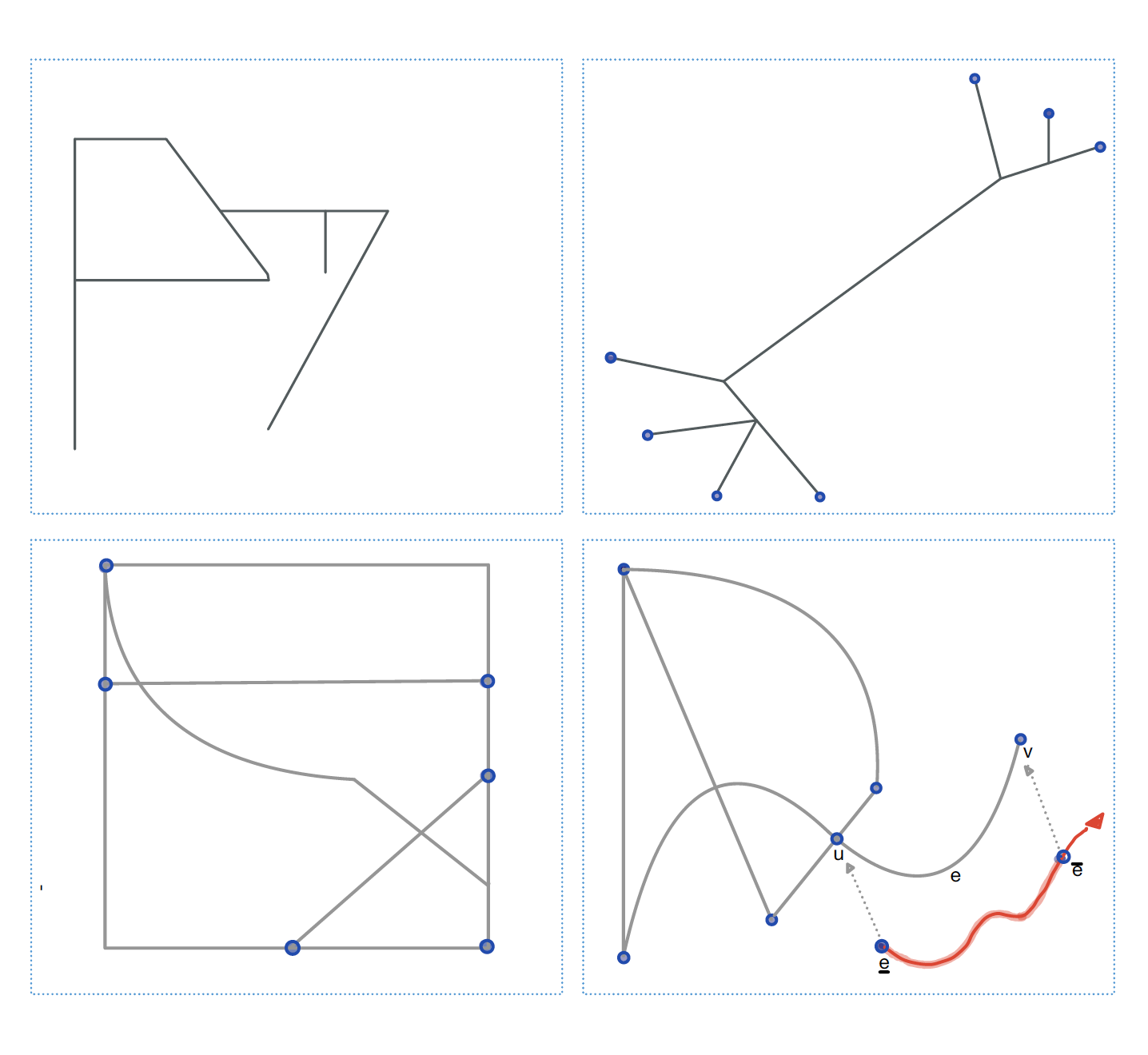}
    \caption{{\em Upper-Left}: an example of linear network;  {\em Upper-Right}: A Euclidean tree with $7$ leaves (blue dots), which may represent a stream network; {\em Bottom-Left}: A graph with Euclidean edges that may represent a road traffic network; crosses between edges with no vertices represent bridges or tunnels. {\em Bottom-Right}: another graph with Euclidean edges; the bijection mapping the vertices $u$ and $v$ into $\underline{e}$ and $\overline{e}$ and the edge $e$ into the open interval $(\underline{e},\overline{e})$ gives an Euclidean system with orientation and a way to measure distances.} \label{basic1}
\end{figure}

A linear network is commonly understood as 
the union of finitely many line segments in the plane, where different edges only possibly intersect with each other at one of their vertices. The upper-left part of Figure \ref{basic1} depicts an abstract drawing for a linear network. More sophisticated pictures are available in the literature, but a simplified version is provided here. Both random fields and point processes over linear networks have been considered in the literature. For continuously indexed random fields (not evolving over time) the reader is referred to \cite{anderes2020}. For point processes over networks, a standard reference is \cite{BADDELEY2021100435}.

A network (or equivalently, a graph), ${\cal G}$, is a pair $({\cal V}, {\cal E})$, with ${\cal V}$ being a collection of nodes (called vertices in graph theory) and ${\cal E}$ denoting a collection of edges. Clearly, a linear network as above is a special case of network.
Generalized networks as defined by \cite{anderes2020} allow for nonlinear edges. The problem with nonlinear edges stands mainly in how to measure distances between any pair of points belonging either to the set of vertices, or to the edges. The problem is solved by \cite{anderes2020} who propose graphs with Euclidean edges: those are sophisticated topological structures that allow {\em distances} in the following way. Each graph has a collection of bijection mappings, such that each edge in the graph is mapped into an open interval (see bottom-right of Figure~\ref{basic1}), and every pair of vertices connected by the edge is mapped into two points at the extremes of the same interval. This creates a Euclidean system with an orientation and a suitable way to measure distances. Rigorous definitions of these topological structures are given in the Section \ref{A}. The bottom-right part of Figure \ref{basic1} depicts a typical graph with Euclidean edges. We note that the edge $e$ is mapped into an open interval, and that the two vertices $(u,v)$ are mapped into the endpoints of the interval, denoted $(\underline{e},\overline{e})$. The path merging $\underline{e}$ and $\overline{e}$ is highlighted in orange in the picture. The geodesic distance is the length of such a path, and is denoted $d_{G}$ throughout the manuscript. More accurately, the geodesic distance, $d_G$, is the length of the shortest path merging any pair of points belonging to ${\cal G}$. Another example of graph with Euclidean edges is provided in the bottom-left part of the same figure.

\begin{figure}[t!]
    \centering
        \includegraphics[scale=0.6]{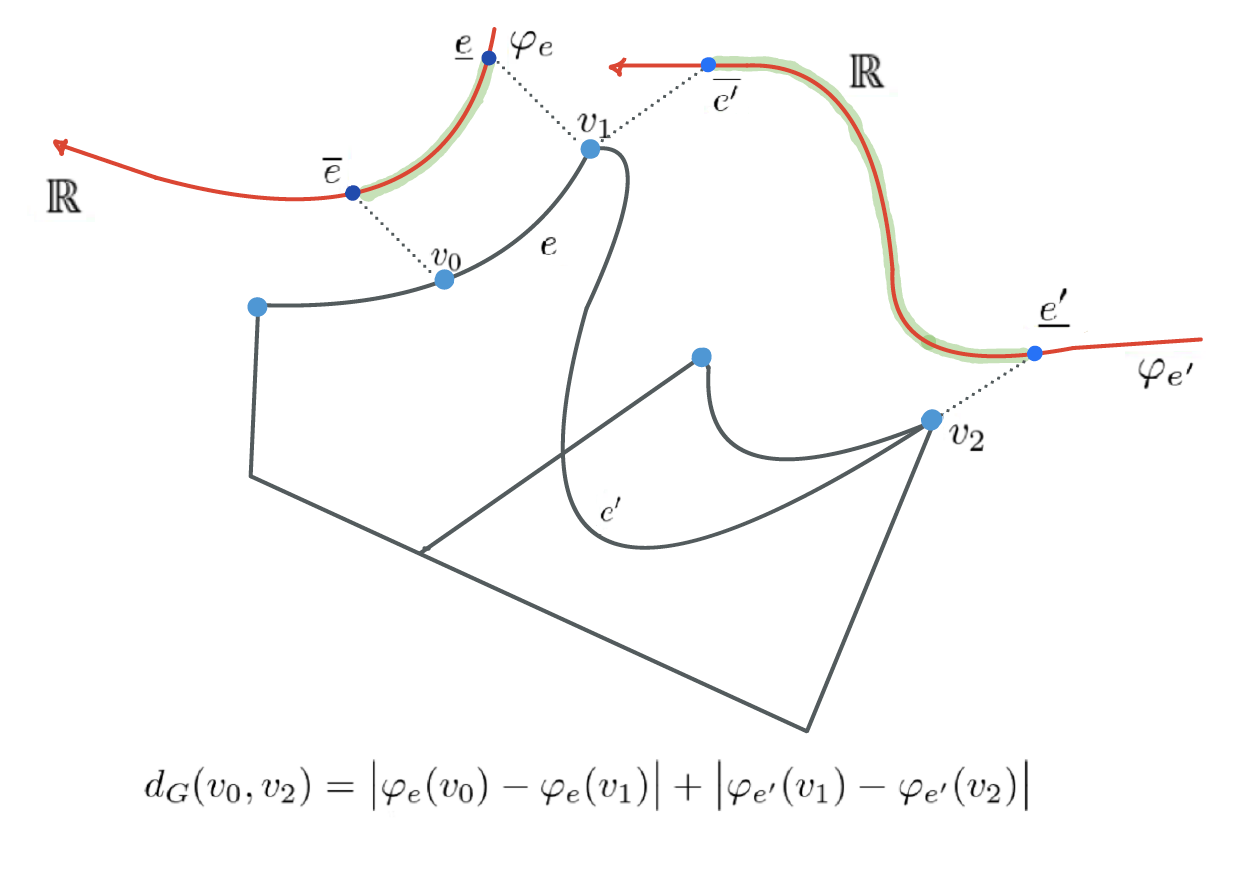}
    \caption{An illustration on how to compute the geodesic distance over a graph with Euclidean edges.} \label{basic2}
\end{figure}

Figure \ref{basic2} allows further illustration of how distances are computed over a graph. We note that to each edge, $e$ and $e'$, are associated two (possibly) different mappings, $\varphi_{e}$ and $\varphi_{e'}$. To calculate the geodesic distance between a vertex, $v_2$ being one extreme of the edge $e'$, and a point $v_0$ lying somewhere on the edge $e$, we sum the length of two paths. The first path is highlighted in green from the red curve. Note:
a) the red curve is a Euclidean coordinate system, with an orientation; 
b) for each edge, $e'$, a bijection $\varphi_{e'}$ is assigned; 
c) to the vertices $v_2$ and $v_1$ we assign respectively the points $\overline{e'}= \varphi_{e'}(v_1)$ and $\underline{e'}=\varphi_{e'}(v_2)$; 
the length of the path is measured through $\big | \varphi_{e'}(v_1)-\varphi_{e'}(v_2) \big|$;  
as a result, we have that 
 $d_G(v_0,v_2)= \big | \varphi_{e}(v_0)-\varphi_{e}(v_1)\big |+\big | \varphi_{e'}(v_1)-\varphi_{e'}(v_2) \big|. $

\begin{figure} 
    \centering
        \includegraphics[scale=0.37]{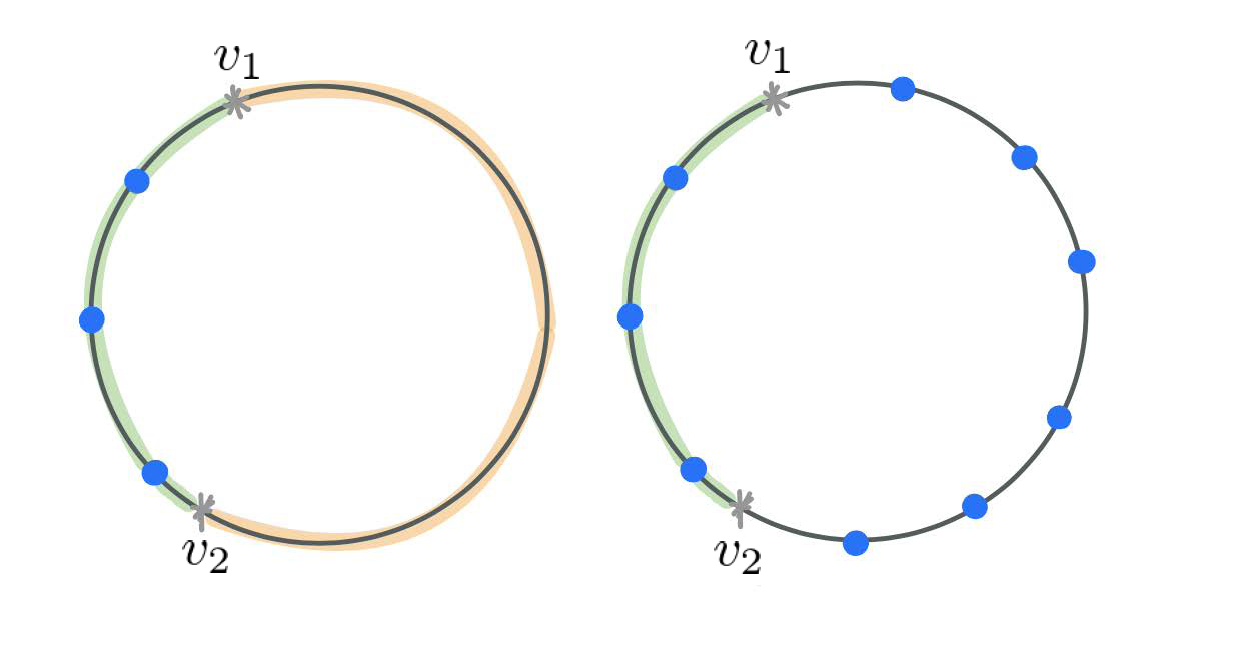}

    \caption{An example of a forbidden (left) and allowed (right) Euclidean cycle.} \label{basic3}
\end{figure}

\cite{anderes2020} provided an accurate description for the technical conditions on a graph to have Euclidean edges. Distance consistency is one of those, and we refer the reader to Figure 1 in \cite{anderes2020} for an illustration. Some graphs are {\em forbidden} with respect to the geodesic distance. Examples are provided by \cite{anderes2020}, and Figure \ref{basic3} shows on the left side an example of a forbidden graph. Distance is inconsistent because the geodesic distance (length of the green arc) is different than the within edge distance (length of the orange arc), which is calculated through $\big | \varphi_{e}(v_1)-\varphi_{e}(v_2)\big |$. On the other hand, the right hand side shows an example where the geodesic and the within edge distances coincide. Such a graph is indeed consistent with respect to the geodesic metric. 

The upper-right part of Figure \ref{basic1} depicts a Euclidean tree, being a special case of a graph with Euclidean edges. It is a tree-like graph (which is planar). Vertices of a Euclidean tree that are connected to one edge only are called leaves. 
As noted in \cite{tang}, an arbitrary point $x$ belongs to ${\cal G}$ when $x \in {\cal V} \cup \bigcup_{e \in {\cal E}} $. As in their paper, we assume that the topological structure of ${\cal G}$ does not evolve over time. 

Our paper considers weakly stationary random fields $\{ Z(\bx,t), (\bx,t) \in {\cal G} \times \mathcal{T} \}$, with ${\cal G}$ being a graph (with Euclidean edges, an Euclidean tree, or a linear network) and $\mathcal{T}$ describing {\em time}, that can be either linear (time is the whole real line, $\mathbb{R}$), or circular (seasonal time on the circle, denoted $\mathbb{S}$). This work focuses on the second order properties of $Z$, with special emphasis on the covariance function, being a linear measure of association between the random variable $Z(\bx,t)$ at point $\bx \in {\cal G}$ an time $t \in \mathcal{T}$, and the random variable $Z(\bx',t')$ at point $\bx' \in {\cal G}$ and time $t' \in \mathcal{T}$. We assume that
\begin{equation}
    \label{G-def}
   {\rm cov} \left \{ Z(\bx,t), Z(\bx',t') \right\} = G \big ( \text{distance}_{\cal G}(\bx,\bx'), \text{separation}_{\mathcal{T}}(t,t') \big), 
\end{equation}
for some suitable function, $G$.
For a random field defined over ${\cal G}$ and not evolving over time, we define the variogram (denoted $\gamma$ throughout) as the variance of the increments of $Z(\bx)$ with respect to $Z(\bx')$, for $\bx,\bx'\in {\cal G}$: 
 $\gamma(\bx,\bx') = \text{var} \left \{ Z(\bx) -Z(\bx')\right \}$.  This also permits an alternative metric, termed {\em resistance} metric in \cite{anderes2020}, and denoted $d_{R}$ throughout. The resistance metric is actually the variogram of a special class of random fields \citep[see][for details]{anderes2020}. The function $G$ will be equipped with either the geodesic or the resistance distance, and with the difference (if time is linear) or the geodesic over the circle (if time is circular).

\section{Main Results: A Non-Mathematical Illustration} \label{sec3a}
\def\bvartheta{\boldsymbol{\vartheta}}

We discuss a general construction here. We consider two parametric classes of functions. Let $p$ and $k$ be two positive integers. Then, we define\\
\def\btheta{\boldsymbol{\theta}}
\begin{equation}
\label{parametric_class} 
{\cal D}_{\btheta}:= \Big \{ \varphi  (x \;  | \btheta ), \quad x \ge 0, \; \btheta \in \R^p \Big \} \qquad \text{and} \qquad {\cal H}_{\bvartheta}:= \Big \{ \psi  (t \; | \bvartheta), \quad t \ge 0, \; \bvartheta \in \R^k \Big \}.
\end{equation}
Clearly, $\btheta$ and $\bvartheta$ are parameter vectors. For instance, the function $\varphi(x \; | \btheta)= \sigma^2 {\rm e}^{-x/a}$ is a suitable example, with $\btheta=(\sigma^2,a)^{\top} \in \R_{+}^2 \subset \R^2$, and with $\top$ denoting the transpose operator. A simple choice from the class ${\cal H}_{\bvartheta}$ is the function 
\begin{equation}
\label{psi_temporal}     
\psi(t \; \big |\bvartheta) = (1+t^a)^b, \qquad t \ge 0,
\end{equation}
with $\bvartheta=(a,b)^{\top}$. Here, $a \in (0,2]$ and $b$ belongs to the interval $(0,1]$. 

Our purpose is to use elements of the classes ${\cal D}_{\btheta}$ and ${\cal H}_{\bvartheta}$ to create a new class of space-time covariance functions with a wealth of practical examples and interactions between space and time. 
 Specifically, we propose the following construction. For two given functions $\varphi$ and $\psi$ belonging to the class ${\cal D}_{\btheta}$ an ${\cal H}_{\bvartheta}$, respectively, we define the function $G_{\alpha,\beta}(\cdot,\cdot \; |\btheta, \bvartheta ): [0,\infty)^2 \to \R$ through
\begin{equation} \label{construction_0}
G_{\alpha,\beta}  (x,t \;  | \btheta, \bvartheta ) = \frac{1}{\psi(t | \bvartheta)^{\alpha}}  \varphi \Bigg ( \frac{x}{\psi(t | \bvartheta)^{\beta}}\; \Big | \btheta \Bigg ), \qquad x,t \ge 0.    
\end{equation}
The parameters $\alpha$ and $\beta$ have been left intentionally outside the vectors $\btheta$ and $\bvartheta$ because of their physical interpretation, as will be clarified subsequently. Hence, the problem is to find conditions on the functions $\varphi$, $\psi$, and on the values of $\alpha$ and $\beta$ such that 
\begin{equation}
    \label{construction}
    {\rm cov} \{ Z(\bx,t), Z(\bx',t') \} = G_{\alpha,\beta} \Big ( d_{\cdot}(\bx,\bx'), |t-t'| \; \Big | \btheta,\bvartheta \Big), \qquad \bx,\bx' \in {\cal G}, \; t,t' \in \R
\end{equation}
is a positive definite function. Here, we use $d_{\cdot}$ whenever we do not wish to specify any choice between $d_G$ and $d_R$. We now clarify the role of the parameters $\alpha$ and $\beta$. When $\alpha$ and $\beta$ are both positive, $G_{\alpha,\beta}$ corresponds to a functional form that was originally proposed by \cite{Gneiting:2002} for {\em space} being the $d$
-dimensional Euclidean space. Several generalizations of this class are summarized in \cite{porcu201930}. When $\beta$ is positive, the spatial distance is rescaled by temporal dependence. When $\beta$ is negative, then the function acting on temporal dependence multiplies the spatial distance. 

The functional and parametric conditions ensuring $G_{\alpha,\beta}$ to become a covariance function are carefully explored and justified in  Section \ref{sec3}. Proofs are technical and require a solid mathematical background, hence we provide a simplified exposition here. For technical details, the reader is referred to Theorems \ref{thm1}, \ref{thm2}, \ref{thm3} and \ref{thm-spectral} in Section \ref{sec3}.

\subsection{Examples from the New Class of Space-Time Covariance Functions} \label{sec3b}

We show here how the class of space-time covariance functions proposed in (\ref{construction}) can be adapted for a wealth of practical situations and different interactions between space and time. To select any example from this class, the practitioner should take into account: 
\begin{itemize}
    \item The reference space: a linear network, Euclidean tree, or graph with Euclidean edges;
    \item The temporal component: time being linear or circular;
    \item The type of spatial distance: the geodesic, $d_G$, or the resistance metric, $d_R$. In turn, this choice depends on the reference space and on the function $\varphi$ used for the construction (\ref{construction_0}). Details are in Theorems \ref{thm1} and \ref{thm3};
    \item The fact that the function $\varphi$ from the class ${\cal D}_{\btheta}$ is strictly positive on the positive real line, or compactly supported. 
\end{itemize}
Some examples follow.

\subsubsection*{Example 1.}

We consider a graph with Euclidean edges, ${\cal G}$, equipped with the resistance metric, $d_R$. We consider the function $x \mapsto \varphi \left  (x \; |\btheta \right ) = \sigma^2 \{ 1+(x/c_{S})^{b_S} \}^{-\delta_{S}}$, for $x \ge 0$. Here, the positive parameter $c_{S}$ rescales spatial distance, while the parameters $b_S \in (0,1]$ and $\delta_S >0$ are related to fractal dimension and long memory of the associated random process. We can use the function $\psi$, with parameter vector $\bvartheta=(a_T,b_T)^\top$ as defined through (\ref{psi_temporal}). Throughout, we fix $b_T=1$ with no loss of generality. Hence, a direct application of Theorem \ref{thm1} ensures that 
\begin{equation}
\label{example1} G_{\alpha,\beta}(d_R(\bx,\bx'), |t-t'| \; \big |\btheta,\bvartheta)= \frac{\sigma^2}{\left  \{1+\left ( \frac{|t-t'|}{c_T} \right )^{a_T} \right \}^{\alpha}} \left ( 1 + \left [ \frac{d_R(\bx,\bx')}{c_S  \left \{1+\left ( \frac{|t-t'|}{c_T} \right )^{a_T} \right \}^{ \beta}}  \right ]^{b_S}     \right )^{-\delta_S},
\end{equation}
for $\bx,\bx' \in {\cal G}$ and $t,t' \in \R$, is a valid covariance functions provided $\alpha >0$ and $\beta \in (0,1]$. 

A similar example can be created by replacing the function $\varphi$ above with the function 
$\varphi(x\; \big |\btheta) = \sigma^2 \{ 1 - x^{b_S \delta_S}(1+ x^{b_S})^{-\delta_S} \}$, $x \ge 0$. Using the function $\psi(t | \bvartheta)=(\eta+ t^{a_T})$, $\eta>0$, and after appropriate rescaling over space and time, we get 
\begin{eqnarray}
\label{example1_b} && G_{\alpha,\beta} (d_R(\bx,\bx'), |t-t'| \; \big | \btheta,\bvartheta) = \frac{\sigma^2}{\left  \{\eta+\left ( \frac{|t-t'|}{c_T} \right )^{a_T} \right \}^{ \alpha}}  \\ &&   \times\left \{  1 - \left [ \frac{d_R(\bx,\bx')}{c_S  \left \{\eta+\left ( \frac{|t-t'|}{c_T} \right )^{a_T} \right \}^{ \beta}} \right ]^{b_S \delta_S}  \left ( 1 +   \left [ \frac{d_R(\bx,\bx')}{c_S  \left \{\eta+\left ( \frac{|t-t'|}{c_T} \right )^{a_T} \right \}^{ \beta}}  \right ]^{b_S}     \right )^{-\delta_S} \right \}, \nonumber
\end{eqnarray}
for $\bx,\bx' \in {\cal G}$ and $t,t' \in \R$, is a valid covariance functions provided $\alpha >0$ and $\beta \in (0,1]$. 

For both examples, the resistance metric might be replaced with the geodesic, and the reader is referred to point 3 in Theorem \ref{thm1} for details. 

\subsubsection*{Example 2.}

Covariance functions with compact support play an important role when the graph is a Euclidean tree with a given number of leaves. Details are provided through Theorem \ref{thm2} in Section \ref{sec3}. An illustration is provided below. 
We consider the function 
$$ \varphi ( x \; | \btheta )= \sigma^2 \left ( 1 - \frac{x}{c_S} \right )_{+}^{\nu_S}, \qquad x \ge 0, 
$$ 
where $c_S>0$, $\nu$ is positive and has a lower bound that is specified through Theorem \ref{thm2}. Here, $(x)_+$ stands for the positive part of the real number $x$.
The parameter $c_S$ determines the support of the function, because $\varphi$ is identically equal to zero whenever $x \ge c_S$. Specifically, we have for example that 
\begin{equation}
\label{example3_c} G_{\alpha,\beta } 
(d_{\cdot}(x,x'), |t-t'| \;\big |\btheta,\bvartheta )= \frac{\sigma^2}{\left  \{1+\left ( \frac{|t-t'|}{c_T} \right )^{a_T} \right \}^{ \alpha}}  \left [ 1 - \frac{d_{\cdot}(x,x')}{c_S \left  \{1+\left ( \frac{|t-t'|}{c_T} \right )^{a_T} \right \}^{ \beta} } \right ]_+^{\nu_s}
\end{equation}
is a valid construction. Indeed, this is possible if the graph ${\cal G}$ is a Euclidean tree with a given number of leaves, $m$. The parameters $\nu_S$ and $\alpha$ depend linearly on $m$ (see Theorem \ref{thm2}), and $\beta$ belongs to the interval $[0,1)$. Also, the geodesic distance can be replaced by the resistance metric with no harm (see Theorem \ref{thm2}). An important feature of the covariance function in Equation (\ref{example3_c}) is that it is dynamically compactly supported. That is, for every fixed time $t,t'$, the function $G_{\alpha,\beta}$ is compactly supported over a ball embedded in ${\cal G}$ with radius $\psi(|t-t'| \; \big | \bvartheta)$. This feature has been well studied in spatial statistics, and the reader is referred to \cite{Porcu-Bevilacqua-Genton-2} for a modeling perspective, as well as to \cite{BFFP} for the implications of using compact support for modeling, estimation and prediction under the so-called infill asymptotic framework \citep{stein-book}.

\subsubsection*{Example 3.}
When the parameter $\beta$ in the covariance $G_{\alpha, \beta}$ is negative, the spatial distance $d_{\cdot}$ is multiplied (no longer rescaled) by temporal dependence. For instance, the function 
\begin{equation}
    \label{example4} G_{\alpha,-1}(d_{\cdot}(\bx,\bx'), |t-t'|\; \big | \btheta,\bvartheta)= \frac{\sigma^2}{\psi(|t-t'| \; \big| \bvartheta)^{\alpha}} \exp \left \{- \frac{d_{\cdot}(\bx,\bx')  }{c_S}\; \psi(|t-t'| \; \big | \bvartheta)\right \}
\end{equation}
is a valid covariance function provide $\alpha \ge 1$ and provided $\psi$ satisfies the conditions in Theorem \ref{thm3}. Many other examples of this kind can be obtained using Theorem \ref{thm3} in concert with suitable choices from Tables \ref{table1} and \ref{table2}.

\subsection{When Time is Circular}

If we suppose time to be circular \citep[as in][]{shirota2017space, white2019nonseparable,mastrantonio}, then the Euclidean distance needs to be replaced by the geodesic distance over the circle. Theorems \ref{thm1}, \ref{thm2} and \ref{thm3} provide technical conditions such that the geodesic distance can be used in the function $\psi(\cdot \; | \bvartheta)$. It is worth mentioning that the geodesic distance over the circle has range $[0,\pi]$, so that the function $\psi$ is restricted to this interval. This is stated accurately in the relevant propositions. 

Theorem \ref{thm-spectral} provides a different construction that is based on half-spectral inversion. We do not enter mathematical details, but note here that such a construction allows for examples that cannot be covered through Theorems \ref{thm1}, \ref{thm2} and \ref{thm3}. In particular, all the examples that have been previously introduced do not allow for negative spatial dependencies. The function 
$$ G_{1,1}(d_R(\bx,\bx'), d_{G}(t,t')\; | \btheta,\bvartheta)=\sigma^2 \left \{ \frac{1-\varepsilon}{  1- \varepsilon \psi(d_{R}(\bx,\bx')\; | \bvartheta) \cos d_{G}(t,t')  } \right \}^{\tau}, $$
for $\bx,\bx'\in \cal G$, $t,t'\in \mathbb{S}$,
is a valid covariance function for any graph with Euclidean edges cross circular time. Here, $\psi$ needs to be a positive definite function over the circle with the additional requirement that $\psi(0\; | \bvartheta) = 1$. The parameter $\varepsilon$ belongs to the open interval $(0,1)$. More examples are reported in Table \ref{tabula_rasa}.

\subsection{Previously Proposed Models}

\cite{menegatto2020gneiting} considered the more general setting of quasi metric spaces and provided sufficient conditions for the structure $G_{\alpha,\beta}(|\cdot|,\sigma(\cdot,\cdot))$ (notice that the arguments are exchanged here) to be positive definite. Here, $\sigma$ is an arbitrary quasi metric. \cite{tang} noticed this fact and considered the pair $({\cal G},d_R)$ as a quasi metric space. As a result, the Menegatto-Porcu-Oliveira construction can be adapted to a covariance function $G_{\alpha,\beta}(|\cdot|,d_R(\cdot,\cdot))$ where the temporal separation is rescaled by spatial dependence. This is unusual in spatial statistics, and for a  constructive criticism the reader is referred to \cite{porcu-alegria-furrer}, with the references therein. 

For the case of Euclidean trees with a given number of leaves, \cite{tang} proposed what they term {\em metric models}. Let $\rho_{1,n}$ be the $\ell_1$ distance in $\R^n$. For a function $C$ such that $C(\rho_{1,n+1}(\cdot,\cdot))$ is positive definite, arguments in Theorem 4 of \cite{anderes2020} show that $C(d_{\cdot}(\cdot,\cdot)+|\cdot|)$ is positive definite over a Euclidean tree cross the real line. The construction is clearly reminiscent of zonal anisotropy in geostatistics, whose adaptation to the space-time setting has been abundantly criticized, and we refer the reader to \cite{chiles}, \cite{Gne:2002b}, \cite{stein-jasa} and \cite{Porcu20111293} amongst others.

\subsection{Related Constructions and Forbidden Models}

\subsubsection*{A partially forbidden model}

The Mat{\'e}rn class of functions has been the cornerstone of spatial statistics for longtime. We refer the reader to \cite{stein-book} and more recently to \cite{bevilacqua2022unifying} for a thorough account. We define it here through
\begin{equation}
\label{matern} {\cal M}_{\nu}(x) = \frac{2^{1-\nu}}{\Gamma(\nu)} x^{\nu} {\cal K}_{\nu}(x), \qquad x \ge 0,
\end{equation}
where ${\cal K}_{\nu}$ is a modified Bessel function of the second kind of order $\nu>0$. The parameter $\nu$ allows to index mean square differentiability for the associated random process. Given the massive use of the Mat{\'e}rn family in spatial statistics, one might be tempted to choose
$$ \varphi(x \; | \btheta) = \sigma^2 {\cal M}_{\nu_S} \left ( \frac{x}{c_S} \right ), \qquad x \ge 0, $$
with $\btheta= (\sigma^2, c_S, \nu_S)^{\top}$, where the three parameters index the variance, the spatial scale, and the smoothness, respectively. A similar calculation as in Example 1 would yield \\
\begin{equation}
\label{example2} G_{\alpha,\beta} (d_R(\bx,\bx'), |t-t'| \; \big | \btheta,\bvartheta )= \frac{\sigma^2}{\left  \{1+\left ( \frac{|t-t'|}{c_T} \right )^{a_T} \right \}^{ \alpha}} {\cal M}_{\nu_S} \left ( \frac{d_R(\bx,\bx')}{c_S \left  \{1+\left ( \frac{|t-t'|}{c_T} \right )^{a_T} \right \}^{ \beta} } \right ).
\end{equation}
Unfortunately, arguments in Theorem 1 in \cite{anderes2020} in concert with Theorem \ref{thm1} in Section \ref{sec3} show that the parameter $\nu_S$ is restricted to the interval $(0,1/2]$ to ensure positive-definiteness. For such an interval, the associated process is continuous but not mean square differentiable. Hence, this covariance is not suitable to index spatial smoothness. 

\subsubsection*{Forbidden models and unclear cases}

For ${\cal G}$ being any general graph with Euclidean edges, Section \ref{sec3} shows that the functions $\varphi$ and $\psi$ involved in the composition $G_{\alpha,\beta}$ as in Equation (\ref{construction_0}) are both strictly positive. While Theorem \ref{thm2} in Section \ref{sec3} shows that covariance functions can attain negative values for a relatively small number of leaves, Corollary 3 in \cite{anderes2020} proves that covariance functions on trees with {\em any} number of leaves must be strictly positive, or identically equal to zero after a given lag. Further, the condition of non negativity is only necessary. Sufficient conditions based on isometric embeddings from the metric space $(\R^n, \rho_{1,n})$ into the quasi metric space $({\cal G},d_{\cdot})$, where the number of leaves is related to the dimension $n$ where the original space is defined can be inferred from \cite{zastavnyi2000positive}.

An important implication for data analysis is that the model in Equation (\ref{example3_c}) is definitely not suitable for Euclidean trees with a large number of leaves. In fact, the condition $\nu \ge 2n -1$ from Theorem \ref{thm2}, with $n= \ceil{m/2}$ and $m$ being the number of leaves, implies that for $m$ large enough the kernel almost vanishes except at the origin. Such an inconvenience is clearly shared by the metric models that have been introduced by \cite{tang} to analyze data over a tree with a given number of leaves. Example \ref{example4} overcomes this inconvenience, as the function $\varphi$ in the composition is not related to the number of leaves. Hence, the example in Equation (\ref{example4}) is more recommendable to deal with Euclidean trees with a large number of leaves.

Finding covariances with negative values on a general graph with Euclidean edges is elusive. Theorem 1 in \cite{anderes2020} provides a sufficient condition, and all the functions satisfying such a condition are strictly positive. It is unclear whether any choice of function $\varphi$ attaining negative values can preserve positive definiteness. \cite{emery-porcu-22blind} show that this is doable when {\em space} is the metric space $(\R^n, \rho_{2,n})$, with $\rho_{2,n}$ denoting the Euclidean distance in $\R^n$. Apparently, the elegant isometric argument in \cite{anderes2020} cannot be used in this case, and this remains an open problem. 

\subsection{New Models Commuted from Old Literature}

The proofs of Theorems \ref{thm1}, \ref{thm2} and \ref{thm3} show that typical scale mixtures arguments can be used to adapt space-time covariance functions that have been proposed for the setting of the metric space $(\R^n \times \R, \rho_{2,n}^2,\rho_{1,n} )$. Here we list the most prominent constructions. 
\begin{enumerate} 
\item The \emph{quasi-arithmetic} class \citep{porcu-mateu-christakos};  \item The scale mixtures as in \cite{fonseca}, \cite{schlather}, and \cite{Apanasovich:Genton:2010}; \item Other scale-mixture based constructions as in \cite{ Porcu:Gregori:Mateu:2006}, \cite{porcu-mateu-bevilacqua}, \cite{porcu-mateu}, and \cite{APFM}. 
\end{enumerate}    
Other popular constructions can be adapted from earlier literature. For instance, \cite{emery-peron-porcu} proposed linear combinations of products of covariance functions defined over graphs with temporal covariance functions. They provided conditions for at least one weight in the linear combination to be negative.

\section{Simulation Study} \label{sec4}

\begin{figure}[H]
    \centering
    \includegraphics[width = 0.48\textwidth]{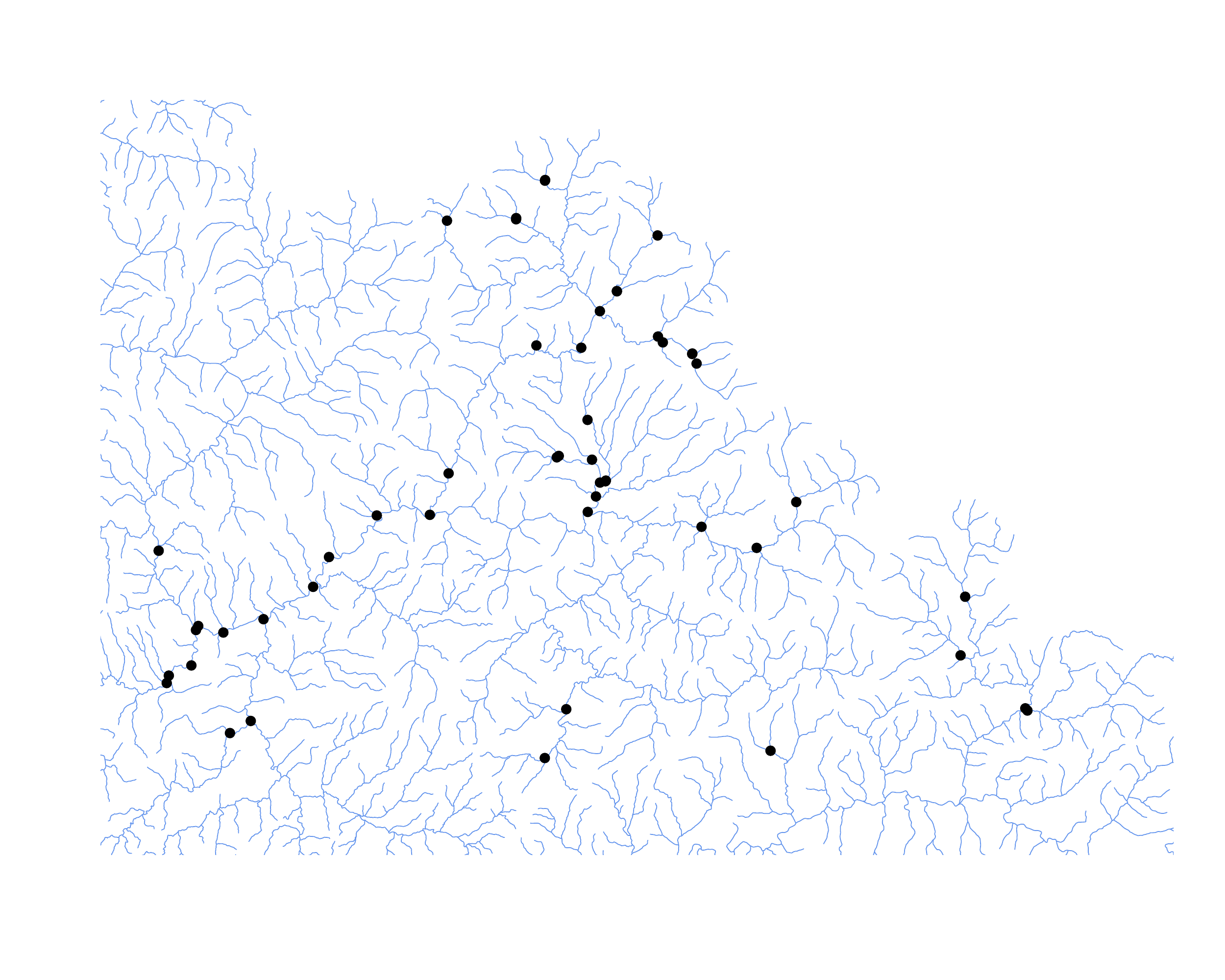}
        \includegraphics[width = 0.48\textwidth]{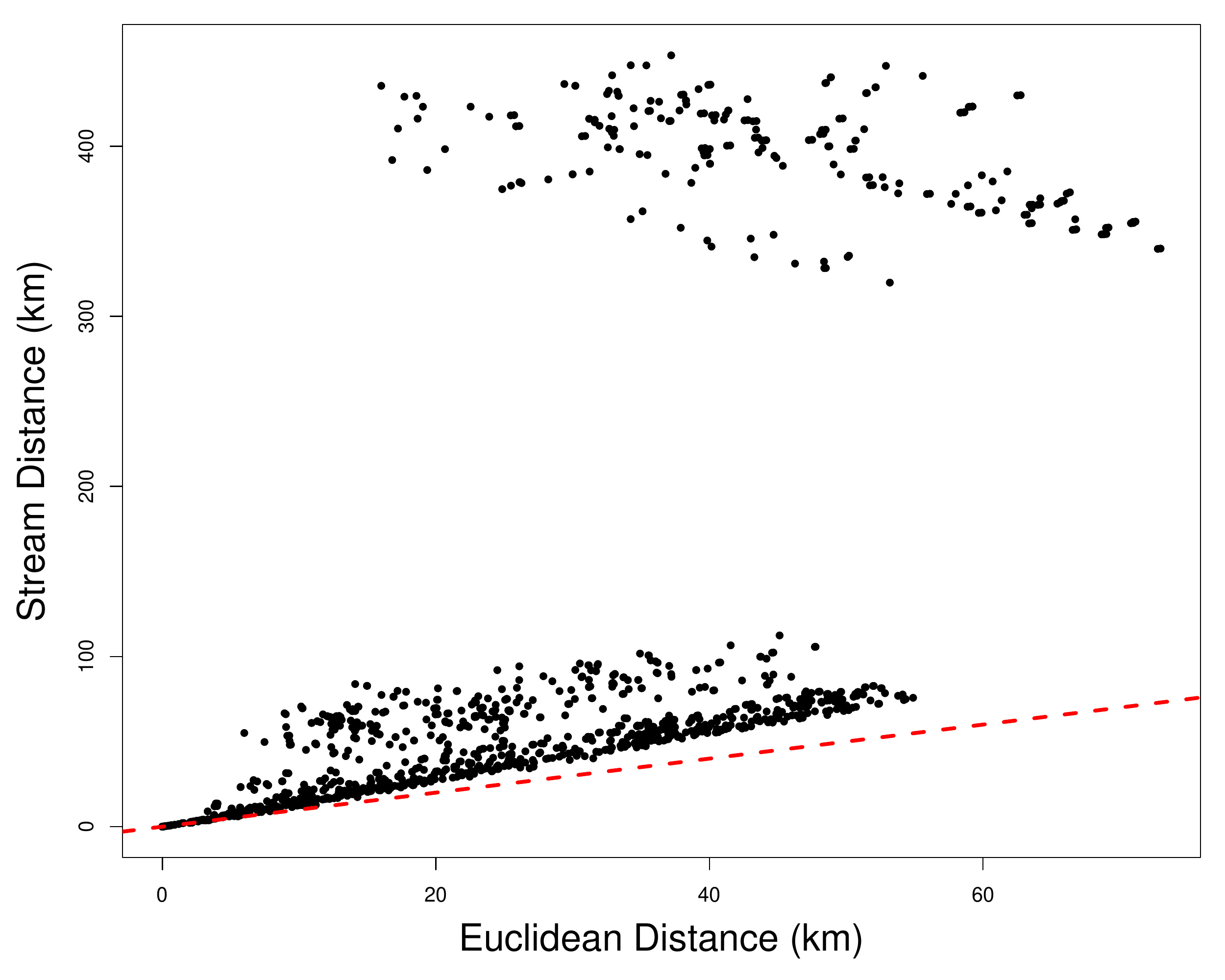}

    \caption{(Left) Locations of sites on river network used for the simulation study. (Right) Comparison of Euclidean distance and network (stream) distance for the sites used in the simulation study. Dashed line has a slope of one and marks equality.}
    \label{fig:sim_data}
\end{figure}

In this simulation study, we consider data simulated over a river network. Specifically, we consider a subset of $n_s= 50$ sites on the Clearwater River Basin in Idaho, USA (See Figure \ref{fig:sim_data}). These locations are derived from data used by \cite{isaak2018principal}, available at \url{https:// www.researchgate.net/publication/325933910_Principal_components_of_thermal_ 30regimes_in_mountain_river_networks}. We calculate the Euclidean distance and plot it against the distance over the river network in Figure \ref{fig:sim_data} to emphasize the difference in distances depending on the metric used. For each location, we simulate 10 random time points, distributed uniformly between 0 and 1, giving 500 data points. Because river networks generally present unique challenges with flow direction and river connections \citep{ver2006spatial}, we emphasize that we only use this network structure as an illustration.

We consider four experiments in this study, and, for each experiment, we simulate $n_{sim} = 1000$ datasets through
\begin{equation}\label{eq:sim_mod}
    \by \sim \text{Normal}\left(\bzero,\bSigma  +  \tau^2 \mathbb{I}\right),
\end{equation}
where $\tau^2 = 0.1$ is a fixed and known nugget effect. Here $\bSigma$ is a covariance matrix with elements determined by Equation (\ref{example1}) which we subsequently denote by $\mathbb{T}$ for the {\em true} covariance model. Specifically, we use the network geodesic distance $d_G$ and fix $\alpha=2$, $a_T=1$, $\delta_S=2$. In the simulation experiments, we treat $c_T$, $c_S$, and $\sigma^2$ as unknown parameters to be estimated. The four simulation experiments differ by varying the spatial range parameter, $c_S = 20, 50, 100, 200$ km, depending on the simulation experiment. For all simulations, we use $c_T = 0.2$ and $\sigma^2 = 0.9$. For every simulated dataset, we fit models using the {\em true} model, as well as two competitors, $\mathbb{C}_i$, $i=1,2$, defined as  
\begin{enumerate}
    \item[($\mathbb{C}_1$)] The same covariance model $\mathbb{T}$, but replacing the network geodesic metric, $d_G$, with the Euclidean distance. 
    \item[($\mathbb{C}_2$)] The covariance model in Equation (\ref{example1_b}), with  $\alpha = 2$, $b_T = 1/4$, $\delta_S = 1/2$, $\eta = 1/2$, and using the geodesic distance $d_G$. As with $\mathbb{T}$, $\sigma^2$, $c_S$, and $c_T$ are unknown.
\end{enumerate}

We examine the results of this simulation study in three ways. To verify that model performance is best for $\mathbb{T}$, in Section \ref{sec:sim_comp}, we present the proportion of times each covariance example had the highest likelihood. To verify that using the correct distance metrics improves parameter estimation, in Section \ref{sec:sim_est}, we compare the parameter estimates under $\mathbb{T}$ and $\mathbb{C}_1$ (these models only differ by the distance metric used for the spatial component). To determine whether we can recapture the true parameters, we compare the estimated model parameters for $\mathbb{T}$ to the true parameters in Section \ref{sec:sim_est}. 


\subsection{Model Comparison}\label{sec:sim_comp}
 
For every simulated dataset, we fit the models that differ in terms of covariance function ($\mathbb{T}$, $\mathbb{C}_1$ or $\mathbb{C}_2$) but have the same model form \eqref{eq:sim_mod}. Using maximum likelihood estimates, we calculate the log-likelihood for each model to identify the covariance example with the highest likelihood. In Table \ref{fig:best_mod}, we present the proportion of simulations where each example had the highest likelihood, calculated for each experiment ($c_S = 20, 50, 100, 200$). Even for relatively short spatial range parameters, the model $\mathbb{T}$ was chosen 92\% of the time; however, $\mathbb{T}$ was chosen even more frequently as the range parameter $c_S$ increases. Thus, we find that as the range parameter increases ({\em i.e.}, the persistence of spatial correlation increases), it was more important to use network distance.
 
\begin{table}[H]
\centering
\begin{tabular}{lrrr}
  \hline
$c_S$ & $\mathbb{T}$ &  $\mathbb{C}_1$ & $\mathbb{C}_2$  \\ 
  \hline
20 km & 0.920 &  0.080 &   0.000 \\ 
50 km & 0.971 &  0.029 &   0.000  \\ 
100 km & 0.989 &  0.011 &   0.000  \\ 
200 km & 0.998 &   0.002 &   0.000  \\ 
   \hline
\end{tabular}
\caption{The proportion of times each covariance example had the highest likelihood for the simulated dataset.}\label{fig:best_mod}
\end{table}

\subsection{Parameter Estimation}\label{sec:sim_est}

\begin{table}[b!]
\centering
\small
\begin{tabular}{|c|l|rrr|rrr|}
  \hline
 & Covariance & \multicolumn{3}{c}{MAE} & \multicolumn{3}{c}{RMSE}   \\
$c_S$  & Example & $\hat{\sigma}^2$ & $\hat{c}_S$ & $\hat{c}_T$ & $\hat{\sigma}^2$ & $\hat{c}_S$ & $\hat{c}_T$ \\ 
  \hline
\multirow{2}{*}{$20$ km}  & $\mathbb{T}$ & \textbf{0.101} & \textbf{3.664} & \textbf{0.013} & \textbf{0.127} & \textbf{4.610} & \textbf{0.017} \\ 
  & $\mathbb{C}_1$ & 0.105 & 7.616 & 0.014 & 0.133 & 8.090 & 0.018 \\ 
    \hline
\multirow{2}{*}{$50$ km} & $\mathbb{T}$ & \textbf{0.131} & \textbf{10.228} & \textbf{0.015} & \textbf{0.166} & \textbf{12.935} & \textbf{0.019} \\ 
 & $\mathbb{C}_1$ & 0.140 & 23.104 & \textbf{0.015} & 0.176 & 24.024 & 0.020 \\ 
    \hline
\multirow{2}{*}{$100$ km}  & $\mathbb{T}$ & \textbf{0.160} & \textbf{21.746} & \textbf{0.015} & \textbf{0.198} & \textbf{28.226} & \textbf{0.019} \\ 
  & $\mathbb{C}_1$ & 0.175 & 53.048 & 0.017 & 0.219 & 54.596 & 0.021 \\ 
    \hline
\multirow{2}{*}{$200$ km} &  $\mathbb{T}$ & \textbf{0.192} & \textbf{51.043} & \textbf{0.017} & \textbf{0.243} & \textbf{66.478} & \textbf{0.021} \\ 
  & $\mathbb{C}_1$ & 0.214 & 119.375 & 0.019 & 0.274 & 122.218 & 0.024 \\ 
   \hline
\end{tabular}
\caption{Average distances between estimated and true parameters under all four covariance functions from the four simulation experiments. Bolded numbers indicate the smallest distance (best performance). ``Exp'' indicates the simulation experiment, ``Model'' refers to the covariance function example used, while MAE and RMSE refer to the absolute and root mean squared error between the estimated and true parameter values.}\label{tab:model_estimation}
\end{table}

In this section, we assess the estimation differences depending on the distance metric used, as well as parameter recovery. As discussed, the data were simulated using $\mathbb{T}$, and we compare the maximum likelihood estimates for $\mathbb{T}$ and $\mathbb{C}_1$ to the true values. For every simulation experiment, we obtain $n_\text{sim}$ sets of estimated parameters. To assess the estimation error between an estimated set of parameters $\hat{\lambda}$ and true values $\lambda$, we use mean absolute error $\text{MAE}(\lambda,\hat{\lambda}) = \frac{1}{n_\text{sim}} \sum^{n_\text{sim}}_{i = 1} |\hat{\lambda}_i  - \lambda| $ and root mean squared error $\text{RMSE}(\lambda,\hat{\lambda}) = \sqrt{\frac{1}{n_\text{sim}} \sum^{n_\text{sim}}_{i = 1} (\hat{\lambda}_i  - \lambda)^2 }.$ We present the simulation errors for $\mathbb{T}$ and $\mathbb{C}_1$ in Table \ref{tab:model_estimation}. In all simulation settings ({\em i.e.}, for all values of $c_S$), MAE and RMSE for all parameters were lower or equal under $\mathbb{T}$, compared to $\mathbb{C}_1$. Unsurprisingly, the largest discrepancies between $\mathbb{T}$ and $\mathbb{C}_1$ are for $c_S$, where the estimation errors for $\mathbb{C}_1$ are nearly twice those for $\mathbb{T}$. The relative difference in estimation errors for $\sigma^2$ and $c_T$ is small.

\begin{figure}[h!]
    \centering
    \includegraphics[width = \textwidth]{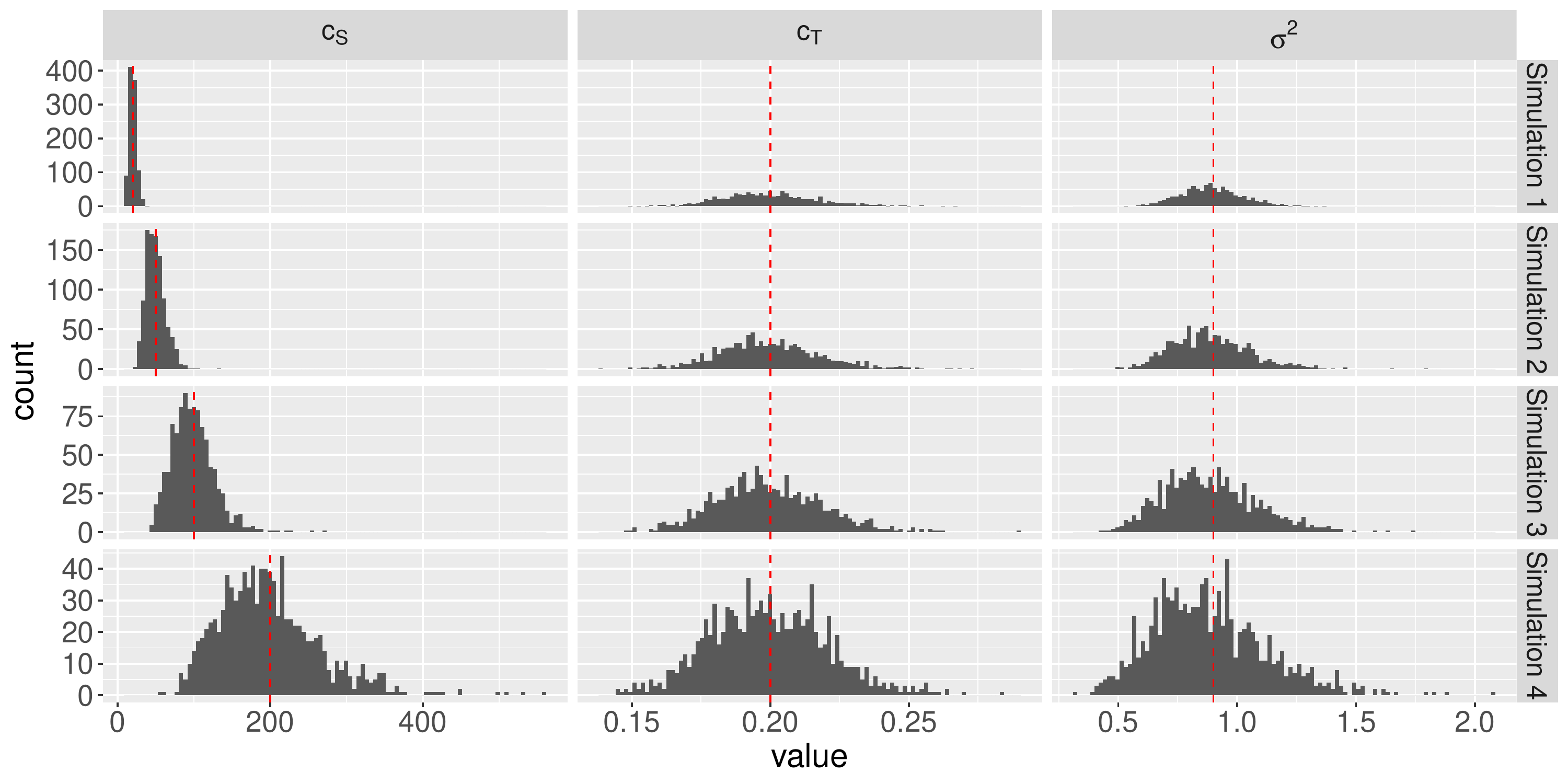}
    \caption{Histogram of model parameters for all simulation experiments. Vertical dashed lines represent the simulation/true values. }
    \label{fig:hist_pars}
\end{figure}

We also plot histograms of estimated parameters under $\mathbb{T}$ against the true values to confirm that the correct parameters can be identified (see Figure \ref{fig:hist_pars}). The true value used to generate the data is well centered in the span of estimated parameters, suggesting that we are effectively able to recover parameters. Because identifiability challenges are common for scale and range parameters for spatial covariance functions \citep{zhang2004inconsistent}, we emphasize that there may still be identifiability challenges that warrant future study.

\section{Data Illustration} \label{sec5}

\begin{figure}[H]
    \centering
    \includegraphics[width = 0.32\textwidth]{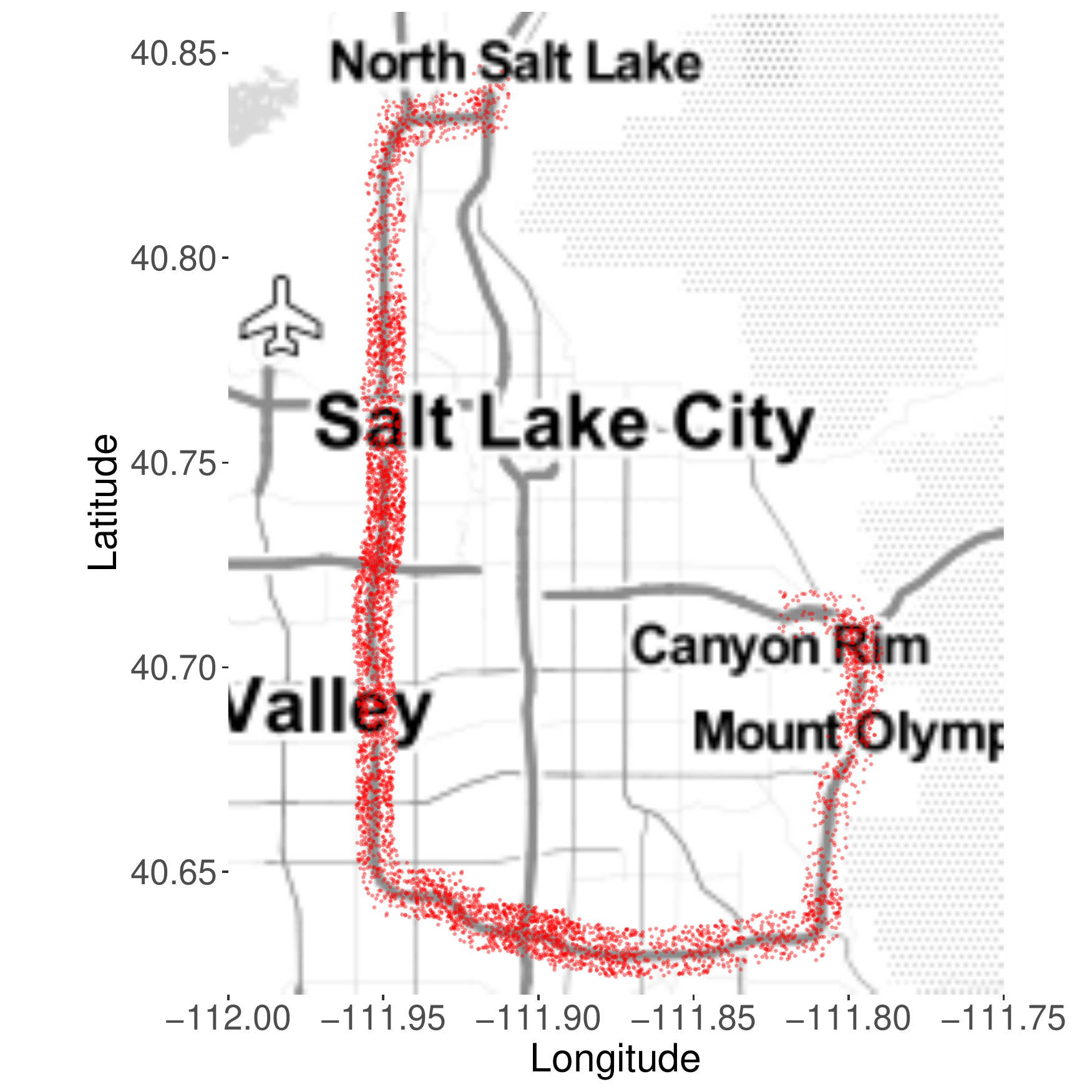}
        \includegraphics[width = 0.32\textwidth]{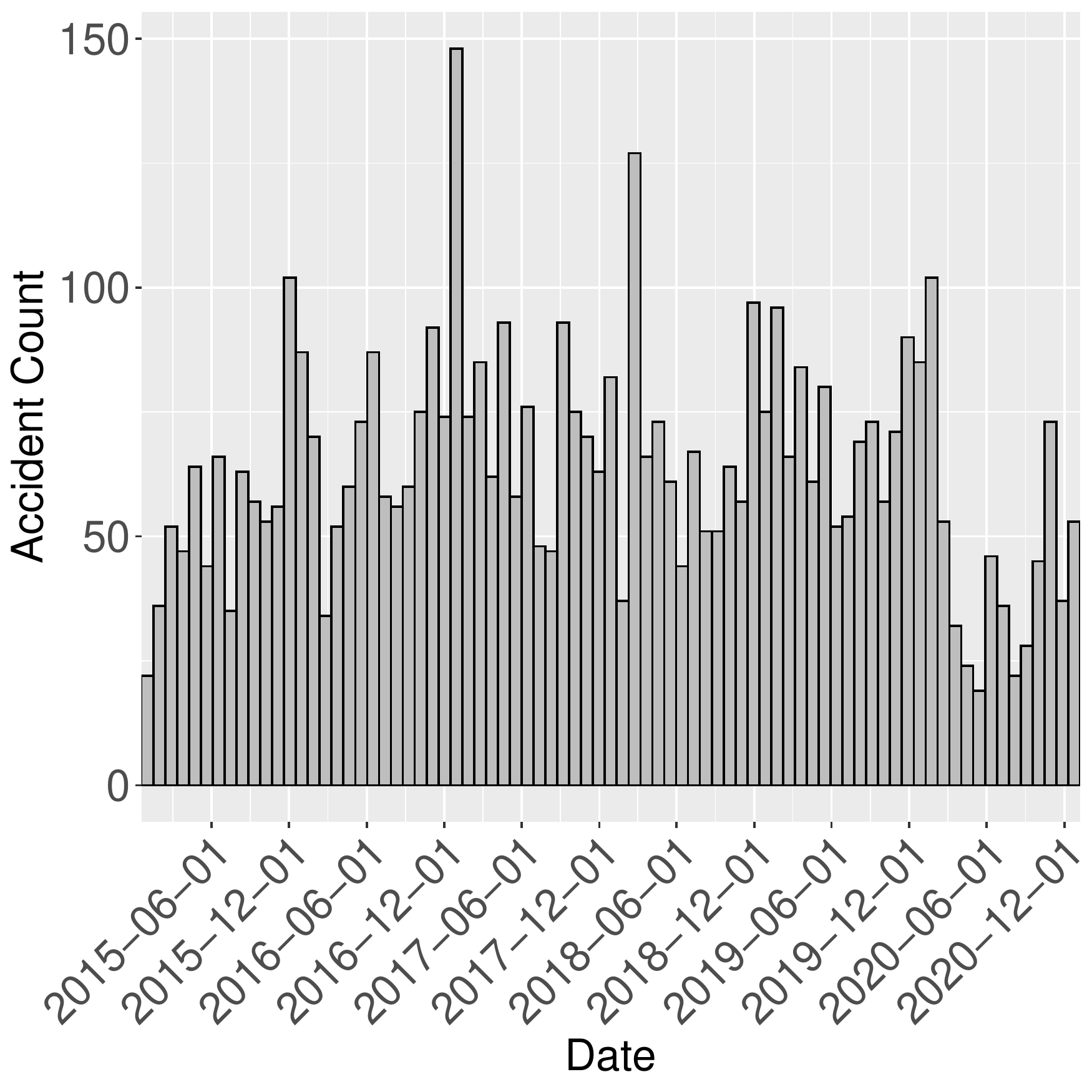}
                \includegraphics[width = 0.32\textwidth]{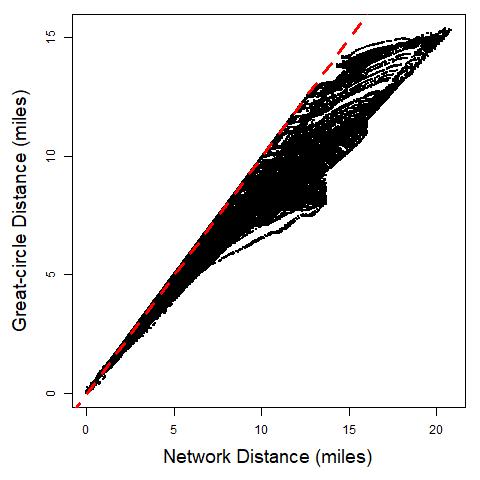}

    \caption{(Left) Jittered accident locations, (Center) a histogram of accident occurrence date, and (Right) a comparison of network and great-circle distance in miles.}
    \label{fig:data_time_loc}
\end{figure}

In this data illustration, we consider traffic accident data from the (approximately) 29-mile I-215 beltway around Salt Lake City, Utah, USA from 2015-2020. Crashes are indexed by time and location (mile post, starting at 0 in the west and terminating near 29 in the north). In total, we observe 5,027 traffic accidents over these six years. Although we are not licensed to share these data publicly, the data can be requested at \url{https://data-uplan.opendata.arcgis.com/}. In Figure \ref{fig:data_time_loc}, we plot jittered locations of these accidents, a histogram of their occurrence date, and comparison of the great-circle distance and network geodesic distance between these crashes. These show spatial and temporal heterogeneity, and significant differences between the network geodesic and great-circle distances. We also point out the drop in accident counts following March 2020 shutdowns due to the COVID-19 pandemic.

In traffic accident modeling, Poisson process models are common \citep[see][for early examples]{jones1991analysis,miaou1993modeling}; however, traffic accident patterns often show overdispersion relative to Poisson processes \citep[see, e.g.,][]{hauer2001overdispersion}. In addition, traffic accident patterns often have higher rates of zeros than Poisson process models support \citep{shankar1997modeling}. Thus, in addition to Poisson models, we also consider Negative Binomial models and Zero-Inflated Poisson models \footnote{We also considered Zero-Inflated Negative Binomial models; however, these models performed poorly.}. To enable simple use of these models and more computationally efficient model fitting, we analyze the data binning over road lengths of 0.5 miles and time windows of approximately one month (30.44 days), giving us counts $y_i$ over $n=4176$ space-time bins. We emphasize, however, that our goal is estimating a continuous space-time intensity surface. In this analysis, we only use one explanatory covariate: the location-specific number of traffic lanes. 

In conjunction with either a Poisson, Negative Binomial, or Zero-Inflated Poisson model, we use models $\mathbb{T}$, $\mathbb{C}_1$ and $\mathbb{C}_2$ from Section \ref{sec4}. We also consider a variation of $\mathbb{C}_2$ using the great-circle distance instead of distance over the road network. We call this choice $\mathbb{C}_3$. Thus, we consider four covariance specifications using different combinations of network/great-circle distances. These covariance structures define Gaussian space-time random effects on the log-mean scale in our models. In total, using three probability mass functions (PMFs) and four covariance functions, we consider $12$ different models.

We fit all models in a Bayesian framework. We use a multivariate log-Normal prior distribution on the covariance parameters $\sigma^2$, $c_S$, and $c_T$, where the location parameters are $1/2$, $1$, and $4$, respectively and the covariance on the log scale is diagonal with elements $1$, $1$, and $1/2$. These prior distributions are informative because of known identifiability challenges with scale and range parameters \citep{zhang2004inconsistent}, but they provide flexibility given the number of crashes, the spatial range of the dataset (in miles), and the time differences in weeks. The intercept and regression coefficient for the number of lanes, which enter the model additively on the log-mean scale, have zero-mean Normal prior distributions with a variance of $100$. For the overdispersion parameter ($r$, normally defined as number of failures) in the Negative Binomial models, we assume that $r \sim \text{Unif}(0,50)$ because the Negative Binomial resembles the Poisson distribution for large $r$. For the Zero-Inflated Poisson distribution, we assume that the zero-inflation probability is Uniform($0,1$) and is constant over space as used by \cite{pew2020justification} on a similar dataset. 

We fit the models using NIMBLE \citep{de2017programming}. We sample the log covariance parameters and regression coefficients separately using blocked multivariate Normal random walks with parameter tuning as described in \cite{shaby2010exploring}. The over-dispersion parameter and zero-inflation parameters are sampled using a Normal random walk. Lastly, we sample the space-time random effects using elliptical slice sampling \citep{murray2010elliptical}. We run this MCMC for $150,000$ iterations, discard a burn-in of $50,000$ iterations, and, for memory reasons, thin the remaining samples to $5,000$ samples.

To compare models, we use the Watanabe-Akaike information criteria (WAIC) \citep{watanabe2010asymptotic}. The WAIC approximates cross-validation and is calculated using the computed log pointwise predictive density $\text{lppd} = \sum^n_{i=1}  \log\{ \frac{1}{M} \sum^M_{m=1} p(y_i| \theta^{m}) \}$, as well as a complexity penalty $p_{WAIC} = \sum^{n}_{i=1} \text{Var}[ \log\{ p(y_i | \theta ) \}]$ \citep[see][for more discussion]{gelman2014understanding}. With these components, WAIC is defined as $-2\text{lppd} +2 p_{WAIC}$, and a smaller WAIC represents a better model. The results of this comparison are given in Table \ref{tab:comp}.


\begin{table}[h!]
\centering
\begin{tabular}{rllrrrr}
  \hline
 & Covariance  & PMF & rWAIC & WAIC & lppd & $p_{WAIC}$\\ 
  \hline
1 & $\mathbb{T}$& Pois & 790.67 & 11307.92 & -4779.60 & 874.36 \\ 
  2 &$\mathbb{T}$ & NB & 799.01 & 11316.26 & -4786.56 & 871.57 \\ 
  3 & $\mathbb{T}$ & ZIP & 774.99 & 11292.25 & -4729.77 & 916.35 \\ 
  4& $\mathbb{C}_1$ & Pois & 742.63 & 11259.88 & -4798.06 & 831.88 \\ 
  5 & $\mathbb{C}_1$ & NB & 880.71 & 11397.97 & -4849.36 & 849.63 \\ 
  6 & $\mathbb{C}_1$ & ZIP & 763.23 & 11280.48 & -4746.37 & 893.87 \\ 
  7 & $\mathbb{C}_2$ & Pois & 554.71 & 11071.96 & -4803.17 & \textbf{732.81} \\ 
  8 & $\mathbb{C}_2$ & NB & \textbf{0.00} & \textbf{10517.25} & \textbf{-4523.12} & 735.51 \\ 
  9 & $\mathbb{C}_2$ & ZIP & 509.45 & 11026.71 & -4717.77 & 795.58 \\ 
  10 & $\mathbb{C}_3$ & Pois & 534.25 & 11051.51 & -4758.27 & 767.48 \\ 
  11 & $\mathbb{C}_3$ & NB & 580.09 & 11097.35 & -4775.35 & 773.32 \\ 
  12 & $\mathbb{C}_3$ & ZIP & 596.64 & 11113.90 & -4770.34 & 786.60 \\ 
  \hline
\end{tabular}
\caption{Model comparison results. ``Pois'' is the Poisson distribution, ``NB'' is the Negative Binomial distribution, and ``ZIP'' is the Zero-Inflated Poisson distribution. We define rWAIC to be relative WAIC (WAIC minus the lowest observed WAIC), so that the best model has a value of 0, and all other models have positive values. }\label{tab:comp}
\end{table}

The results in Table \ref{tab:comp} suggest that the form of the covariance function is important in model performance, as the $\mathbb{C}_2/\mathbb{C}_3$ covariance functions were better than the $\mathbb{T}/\mathbb{C}_1$ alternatives.  The PMF used was also important in model performance, but the best PMF differed depending on the covariance function and distance metric combination. On the whole, network-based distance metrics had lower WAIC; however, there are exceptions. Ultimately, the best model in terms of fit (lppd) and WAIC is the Negative Binomial model with random effects using $\mathbb{C}_2$. We highlight that this covariance function uses the network geodesic to specify space-time random effects and outperforms all models that use the great-circle distance. We interpret the results based on this model.

\begin{figure}[H]
    \centering
        \includegraphics[width = 0.48 \textwidth]{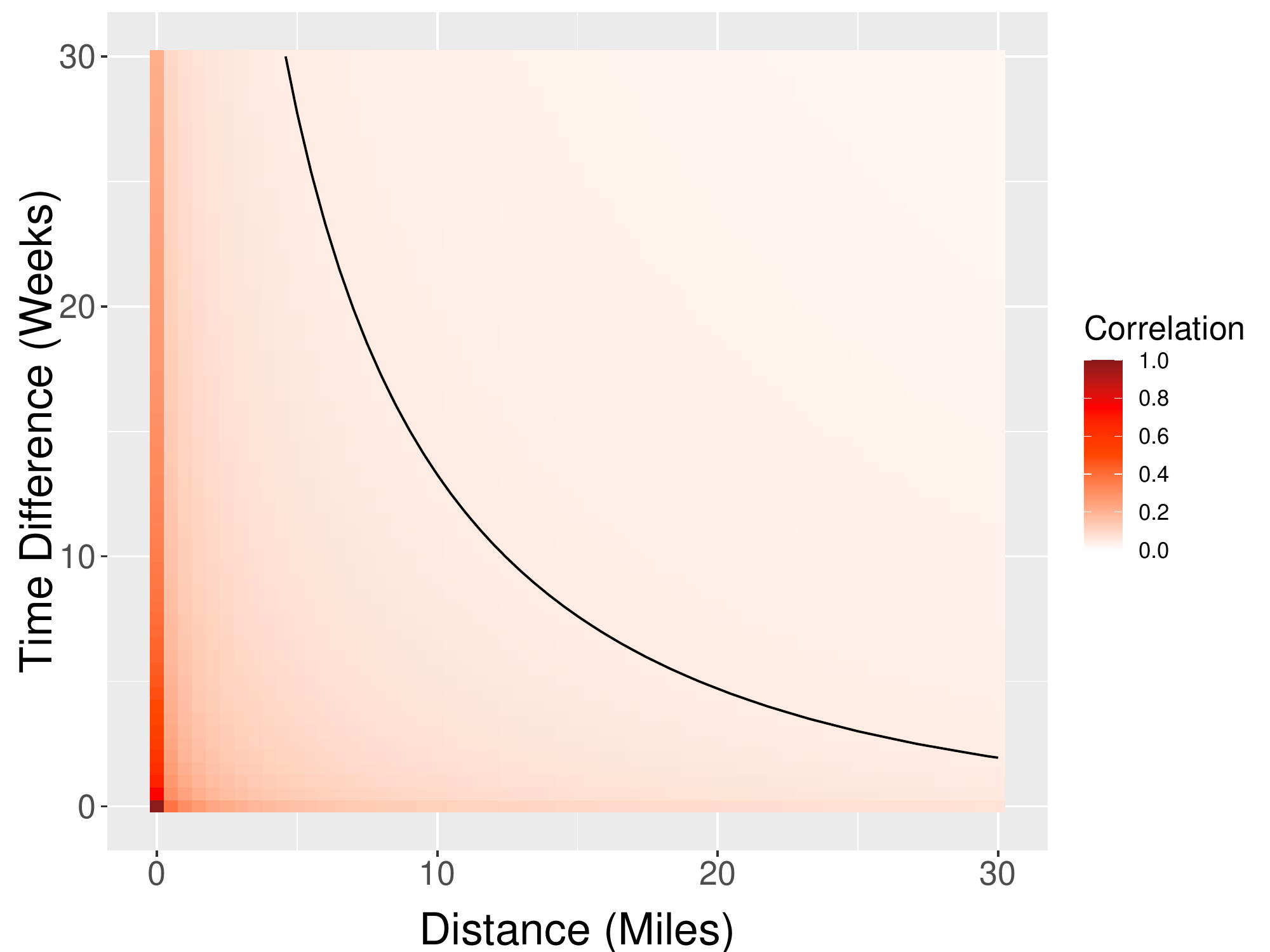}
    \includegraphics[width = 0.48 \textwidth]{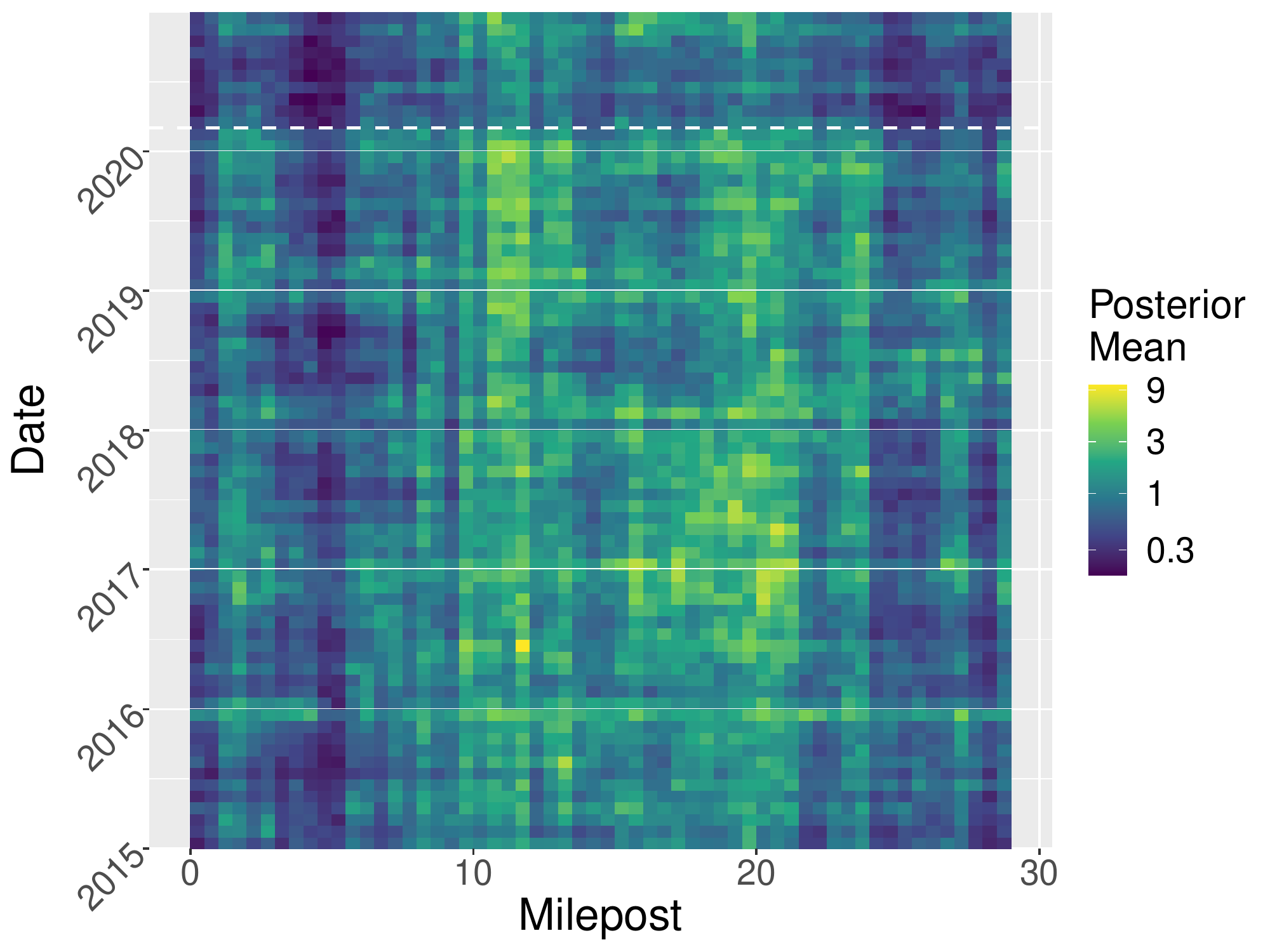}
    \caption{Posterior mean for the (Left) Correlation function over network distance and difference in time (weeks) and (Right) intensity function or average number of crashes per mile$\times$week, where the color scale varies logarithmically. Solid horizontal lines indicate new years, while the thicker dashed line represents the timing of shutdowns due to COVID-19.}
    \label{fig:rand effect}
\end{figure}

We plot the posterior mean of the covariance function for the random effects as a function of network distance and time difference (in weeks) in Figure \ref{fig:rand effect}. In this plot, we include a contour line marking the effective range (the distance/time difference where the correlation reaches 0.05). Although the correlation decays quickly as a function of distance over the network, correlation persists for many weeks over short distances. Similarly, for short time differences (e.g., around five weeks) the posterior mean correlation remains above 0.1 for about five miles and above 0.05 for about 20 miles. For simultaneous points in time, correlations remain about 0.1 for over 15 miles, and the effective range is not reached for any distances observed in this dataset. 

In Figure \ref{fig:rand effect}, we plot the posterior mean of the intensity surface (expected number of crashes per  mile$\times$week) to explore the spatiottemporal patterns in the data. These patterns reveal high spatio-temporal variability in the expected number of accidents.
The COVID-19-related shutdowns in March of 2020 in the United States are some of the most interesting external factors in the time span of these data. The dashed line in the figure indicates when lock downs went into effect. The before/after pattern is captured very clearly along the entire belt route. Along all of I-215, the accident intensity drops rapidly beyond the dashed line, coinciding with decreases in daily commuting. The largest difference in the space-time random effect is near milepost 12, the junction between I-215 and I-15 (the primary highway in Utah). Overall, these results show strong space-time patterns in the data.

In this data illustration, we presented an analysis of traffic accident data on I-215 in Utah, USA. Comparing various models that differed in terms of covariance function, distance metric, and probability mass function, we selected a Negative Binomial model with random effects with covariance function $\mathbb{C}_2$ that uses the network geodesic. Using this model, we explored the estimated space-time random effect and associated correlation structure.

\section{Conclusions} \label{sec6}
We have provided flexible classes of space-time covariance functions that can be used over linear or non linear networks. Our exposition strategy has been devoted to simple illustrations that avoids mathematical details (provided in the Supplementary Material). This allows the practitioner to understand how to use the new models. We have also focused on how to practically calculate distances over networks. Our simulation study highlighted that if the network geodesic is used to generate the data, then model performance and parameter estimation are better using the network geodesic distance than Euclidean distance. In addition, we found that we can effectively recover the parameters of the true covariance function. From our analysis of traffic accident patterns on a simple road network, we found that our best model used the network geodesic distance and outperformed all models that failed to account for network structure. 
This work lays the foundation for many challenges from both theoretical and applied standpoints, among which:
\begin{itemize}
\item[1)] The problem of multivariate covariance functions over networks has, to our knowledge, not been addressed so far by earlier literature. Modern datasets are often characterized by several georeferenced variables that are observed over time. For them, addressing the cross-correlation is of fundamental importance for modeling, estimation, and prediction \citep{Genton:Kleiber:2014}. 
\item[2)] Datasets over linear networks often exhibit nonstationarities over space and time, so that using a covariance function that solely depends on distances might result in unrealistic assumptions. The literature on this subject is elusive so far, and it is unclear how to adapt existing approaches to nonstationarity \citep{paciorek, porcu-mateu-christakos} that have been proposed in Euclidean spaces. In turn, nonstationary models would be the key to a fertile literature on reducibility approaches that allow to interpret a nonstationary random field as a stationary one if commuted into some suitable manifold \citep{porcu2020reduction}.
\end{itemize}

\section*{Acknowledgments}

We acknowledge Jun Tang for sharing code and data. This work was supported by the Khalifa University of Science and Technology Award No. FSU-2021-016 (E. Porcu), NSF-DMS CDS\&E grant 2053188 (P. White), and King Abdullah University of Science and Technology (M. Genton).

\appendix
\section{Background Material} \label{A}

\subsection{Graphs with Euclidean edges}

A network (or equivalently, a graph), ${\cal G}$, is a pair $({\cal V}, {\cal E})$, with ${\cal V}$ being a collection of nodes (called vertices in graph theory) and ${\cal E}$ denoting a collection of edges. ${\cal G}$ is linear if it is the union of finitely many line segments in the plane, where different edges only possibly intersect with each other at one of their vertices. 

\cite{anderes2020} propose graphs with Euclidean edges as a generalization of linear networks. That is, they consider graphs where each edge is associated with an abstract set that is in bijective correspondence with a segment of the real line. This allows to associate each edge with a Cartesian coordinate system to measure distances between any two points located over the edge. Specifically, 
a graph with Euclidean edges is a triple $( {\cal{G}}, {\cal{V}}, \{\varphi_{e}\}_{e \in {\cal{E}} })$ such that:
\begin{itemize}
\item[a)] $({\cal V, cal E})$  is a finite simple connected graph, meaning that the vertex set ${\cal V}$ is finite, the graph has no repeated edges or edge which joins a vertex to itself, and every pair of vertices is connected by a path. 
\item[b)] Each edge $e \in \cal E$ is associated with an abstract set, denoted with the same symbol $e$, 
where the vertex set $\cal V$ and all the edge sets $e \in \cal E$ are mutually disjoint. 
\item[c)] For each edge $e \in \cal E$ and every pair of vertices $(u,v) \in {\cal V} \times {\cal V}$ that is connected by $e$, the mapping $\varphi_e$ is a bijection that applies to $e,u$ and $v$ as follows: $\varphi_e$ maps $e$ into an open interval $(\underline{e}, \overline{e}) \subset{\mathbb{R}}$, and $\varphi_e$ maps $\{u,v\}$ into $\{\underline{e}, \overline{e}\}$. 
\item[d)] Denote with $d_G (u, v) :{\cal V} \times {\cal V}   \to [0,\infty) $ the standard shortest-path weighted graph metric on the vertices of $({\cal V, \cal E})$ with edge weights given by $\overline{e}-\underline{e}$ for every $e \in \cal E$. Then, $d_{G}(u,v)= \overline{e}-\underline{e}$ for all $u,v \in \cal V$ and for each $e \in \cal E$. 
\end{itemize}
The graph ${\cal G}$ endowed with a quasi-distance becomes a quasi-metric space. \cite{anderes2020} propose two alternative metrics. The geodesic distance, $d_G$, is the shortest path merging any pair of points over ${\cal G}$. The resistance metric, $d_R$, is defined as the variance of the increments - {\em i.e.}, a variogram - of a special class of random processes \citep[see][for a detailed {\em essay}]{anderes2020}. Depending on the characteristics of the graph ${\cal G}$, one metric might be used or not to build positive definite functions. When the metrics $d_R$ and $d_G$ can be equivalently used, we use the notation $d_{.}$; this notation slightly deviates from that of \cite{anderes2020} and \cite{tang}. 

Finally, we call Euclidean tree any tree-like graph (which is planar). Vertices of a Euclidean tree that are connected to one edge only are called leaves.

\subsection{Special classes of functions} 
 
We introduce some classes of continuous functions, defined on the positive real line, that will be useful for the construction of parametric classes of space-time covariance functions. 
A function $f : [0,\infty) \to \R_+$ is called completely monotonic if it is continuous,  infinitely differentiable on $(0,\infty)$, satisfying $(-1)^n f^{(n)}(t) \ge 0$, $n \in \mathbb{N}$. Here, $f^{(n)}$ denotes the $n$th derivative and we use $f^{(0)}$ for $f$, where $f(0)$ is required to be finite.  

The Mat{\'e}rn function, ${\cal M}_{\nu}$, as defined through Equation (\ref{matern}), is completely monotonic for $0<\nu \le 1/2   $, and ${\cal M}_{\nu}(\sqrt{\cdot})$ is completely monotonic for any positive $\nu$. 

A function $f:[0,\infty) \to \mathbb{R}$ is called a Stieltjes function if 
\begin{equation}
\label{stieltjes} f(t) = \int_{[0,\infty)} \frac{{\mu} ({\rm d} \xi)}{t+\xi}, \qquad t \ge0,
\end{equation}
where $\mu$ is a positive and bounded measure. We require throughout $f(0)=1$, which implies that 
$ \int \xi^{-1} \mu ({\rm d } \xi) =1$. Let us call ${\cal S}$ the set of Stieltjes functions. It has been proved that ${\cal S}$ is a convex cone \citep{berg2008stieltjes}, with the inclusion relation ${\cal S} \subset {\cal C}$, where ${\cal C}$ is the set of completely monotone functions. The relation \eqref{stieltjes} shows that the function $f(t)= 1/(1+t)$, $t \ge 0$, is a Stieltjes function. Using the fact that $f \in {\cal S}$ if and only if  $1/f$ is a completely Bernstein function \citep[for a definition, see][]{porcu-schilling}, we can get a wealth of examples of Stieltjes functions, as the book by \cite{SSV} provides an entire catalogue of completely Bernstein functions. 

For what follows, we introduce the Askey function, ${\cal A}_{\nu}$, defined as 
\begin{equation}
    \label{askey}
    {\cal A}_{\nu}(t) = \left ( 1 - t \right )_{+}^{\nu}, \qquad t \ge 0, 
\end{equation}
with $\nu>0$ a positive shape parameter. Here, $(x)_+$ denotes the positive part of the real number $x$. 

Let $m$ be a positive integer. A function $f:[0,\infty)$ is called $m$-times monotonic (or multiply monotonic of order $m$) if and only if $(-1)^{m-1}f^{(m-1)}(\cdot)$ is nonnegative, decreasing and convex. This happens if and only if \citep[][with the references therein]{porcu2014generalized}
\begin{equation}
    \label{multiply-monotonic}
    f(t) = \int_{[0,\infty) }  {\cal A}_{m}(st) \mu({\rm d} s), \qquad t \ge 0,
\end{equation}
for $\mu$ positive and bounded measure.

\section{Main Results} \label{sec3}

For simplicity we report again the expression of the class $G_{\alpha,\beta}$ as in Equation (\ref{construction_0}).
\begin{equation*}
G_{\alpha,\beta} (x,t \; | \btheta, \bvartheta ) = \frac{1}{\psi(t | \bvartheta)^{\alpha}}  \varphi \Big ( \frac{x}{\psi(t | \bvartheta)^{\beta}}\; \Big | \btheta \Big ), \qquad x,t \ge 0.    
\end{equation*}
The conditions for positive definiteness, the parametric classes ${\cal D}_{\btheta}$ and ${\cal H}_{\bvartheta}$, and the proof techniques, change substantially depending on the fact that $\alpha$ and  $\beta$ are positive or not. Hence, we provide a separate proof for each case below.

\subsection{The class $G_{\alpha,\beta}$ for $\alpha$ and $\beta$ positive}

For such a case, we provide the following criterion for positive definiteness. 

\begin{theorem} \label{thm1}
Let ${\cal G}$ be a graph with Euclidean edges. Let $G_{\alpha,\beta}$ be the mapping defined through Equation (\ref{construction_0}). Let $\varphi(\cdot \; | \btheta)$ be a parametric family of Stieltjes functions.
 Let $\alpha \ge 1$ and $\beta \in (0,1]$. Then,
\begin{enumerate}
\item if time is linear $(\mathbb{T}=\mathbb{R})$,  $G_{\alpha,\beta}(d_{R}(\cdot,\cdot), |\cdot|^2)$ is positive definite provided $\psi(\cdot\; | \bvartheta)$ is a parametric family of Bernstein functions; 
\item if time is circular $(\mathbb{T}=\S)$, $G_{\alpha,\beta}(d_{R}(\cdot,\cdot), d_{G}(\cdot,\cdot))$ is positive definite provided $\psi(\cdot\; | \bvartheta)$ is the restriction to the interval $[0,\pi]$ of a parametric family of Bernstein functions;
\item if ${\cal G}$ is a graph with Euclidean edges that forms a finite sequential $1$-sum of Euclidean cycles and trees, then for both cases above the resistance metric, $d_R$, can be replaced by the geodesic distance, $d_G$.
\end{enumerate}
\end{theorem}

\begin{lemma} \label{lema1}
Let ${\cal G}$ be a graphs with Euclidean edges. Let $f(\cdot \; |\btheta): [0,\infty) \to \mathbb{R}$ be continuous, with $f(0 \; |  \btheta)<\infty$. Let $g(\cdot \;  |\xi, \bvartheta): [0,\infty) \to \mathbb{R}$ be continuous, with $g(0 \; | \xi, \bvartheta)<\infty$. For a measure space $(X, {\cal F}, \mu)$, with $\mu$ positive and finite, $X=[0,\infty)$, and ${\cal F}$ a sigma-algebra, call 
$F(d_{\cdot},|\cdot| \; \Big | \btheta,\bvartheta)= \int_{X} f(\xi d_{\cdot}(\cdot,\cdot); \; | \btheta) g(|\cdot| \; | \xi, \bvartheta) \mu({\rm d} \xi). $
If:
\begin{enumerate}
    \item $f(\cdot \; |\btheta)$ is completely monotonic; 
    \item $g \left (|\cdot|^2 \; \Big |\xi, \bvartheta \right )$ is positive definite on the real line for all $\xi \ge 0$; 
    \item the mapping $[0,\infty) \ni \xi \mapsto f( \xi d_{\cdot,\cdot} \; | \btheta)g \left (|\cdot|^2 \; \Big |\xi, \bvartheta \right )$ belongs to $L_1(X,{\cal F}, \mu)$ for all $d_{\cdot}$, $|\cdot|$;
\end{enumerate}
then, 
 $F$ is positive definite on the product space $\left (  {\cal G} \times \mathbb{R} \right )^2$.
\end{lemma}
\begin{proof}
We start by noting that $f(\cdot \; |\btheta)$ is completely monotonic if and only if \citep{berg2008stieltjes} 
\begin{equation}
    \label{comp_mon} 
 f(x \; |\btheta) = \int_{[0,\infty)} {\rm e}^{-rx} F({\rm d} r), \qquad x \ge 0,     
\end{equation}
for $F$ positive and bounded. An elementary change of variable in the above integral representation shows that $f(\cdot \; |\btheta)$ being completely monotonic implies $f(\xi \cdot \; |\btheta)$ being completely monotonic for all $\xi \ge 0$. Hence, we can invoke Theorem 1 in \cite{anderes2020} to claim that $f(\xi d_R(\cdot,\cdot) \; |\btheta)$ is positive definite on every graph with Euclidean edges. According to the same result, we have that $f(d_{G}(\cdot,\cdot) \; |\btheta)$ is positive definite provided ${\cal G}$ is a graph with Euclidean edges that forms a finite sequential $1$-sum of Euclidean cycles and trees. Straightforward arguments allow to verify that the mapping 
$$ [0,\infty) \ni \xi \mapsto f( \xi d_{\cdot}{(\cdot,\cdot)} \; | \btheta)g(\xi |\cdot| \; |\xi, \bvartheta) $$
is continuous and bounded. Condition {\em 3}. ensures the scale mixture $F$ to be well defined.  We can then invoke Schur's theorem to claim that such a product provides a positive definite function over the product space $({\cal G} \times \mathbb{R})^2$. 
Positive definite functions are a convex cone that is closed under the topology of finite measures.  Hence, the scale mixture $F$ provides a positive definite function. The proof is completed.
\end{proof}

\begin{proof}[Proof of Theorem \ref{thm1}.]
We provide a constructive argument on the basis of Lemma \ref{lema1} under a specific choice of the functions $f$ and $g$ therein.
We start by noting that
\begin{equation}
    \label{ello} 
     \int_{0}^{\infty} {\rm e}^{-\xi x } {\rm e}^{- r \xi} {\rm d} \xi = \frac{1}{x+r}, \qquad r>0, \; x \ge 0. 
\end{equation}

 We now consider the function $x \mapsto f(x \; | \btheta)= \sigma^2 \exp(-x/c_S)$, for $\sigma^2$ and $c_S$ strictly positive. Clearly, the function $0 \le x \mapsto \exp(-x)$ is completely monotonic. Hence,   $f$ satisfies Condition {\em 1}. in Lemma \ref{lema1}. As for the function $g$, we consider 
 $$ g\left (|\cdot| \; \Big | \xi, \bvartheta \right ) = {\rm e}^{- \xi \psi \left (|\cdot|^2 \; \big | \bvartheta \right )}, $$
 for $\psi$ a Bernstein function. The composition of the negative exponential with a Bernstein function provides a completely monotone function \citep{berg2008stieltjes}. Hence, we can invoke Schoenberg theorem \citep{schoenberg2} to claim that $g$ satisfies Condition {\em 2}. in Lemma \ref{lema1}. Apparently, Condition {\em 3}. is satisfied as well. 
 We now invoke (\ref{ello}) to derive
 $$ \int_{0}^{\infty} \sigma^2 {\rm e}^{-\xi d_R/c_S } {\rm e}^{-  \xi r \psi\left  (|\cdot|^2 \; \big | \bvartheta \right )} {\rm d} \xi = \frac{\sigma^2}{\psi\left (|\cdot|^2 \; \big | \bvartheta \right ) }  \left ( r +\frac{d_R}{c_S \psi \left (|\cdot|^2 \; \big | \bvartheta \right )}\right )^{-1} . \qquad  $$
Assertion {\em 1}. is proved by noting the integral representation in the definition of a Stieltjies function. 
Assertion {\em 2}. is proved similarly, but noting that, for a Bernstein function $\psi$, the composition $\exp(-r \psi\left (\cdot \; \big |\bvartheta \right ) )$ is completely monotonic on the positive real line \citep{berg2008stieltjes}. Hence, the restriction of the Bernstein function, $\psi_{[0,\pi]}$, composed with the negative exponential $\exp(-\cdot)$, provides a restriction of a completely monotone function to the interval $[0,\pi]$. This shows, according to Theorem 1 in \cite{schoenberg1942}, that $d_G \mapsto \exp (- r d_G)$ is positive definite over the circle. The proof is completed using similar arguments as in Assertion {\em 1}.  
\end{proof}

\subsection{The class $G_{\alpha,\beta}$ for $\alpha$ negative and $\beta$ positive}
We start with a technical lemma.

\begin{lemma} \label{lema2}
Let $m$ be a positive integer and denote by $\ceil{\cdot}$ the ceiling function. Let ${\cal G}$ be a Euclidean tree with $\ceil{m/2} \ge 3$ leaves. Let the functions $f$ and $g$, the measure space $(X, {\cal F}, \mu)$, and the function $F$, as being specified through Lemma \ref{lema1}. Let $\rho_{1,m}$ denote the $\ell_1$ norm in $\mathbb{R}^m$.
If:
\begin{enumerate}
    \item $f(\rho_{1,m} \; |\btheta)$ is positive definite;
    \item $g(|\cdot|^2 \; \Big |\xi, \bvartheta)$ is positive definite on the real line for all $\xi \ge 0$; 
    \item the mapping $[0,\infty) \ni \xi \mapsto f( \xi d_{\cdot} \; | \btheta)g(|\cdot|^2 \; |\xi, \bvartheta)$ belongs to $L_1(X,{\cal F}, \mu)$ for all $d_{\cdot}$, $|\cdot|$;
\end{enumerate}
then, 
 $F$ is positive definite on the product space $\left (  {\cal G} \times \mathbb{R} \right )^2$.
\end{lemma}

\begin{proof}
The proof is obtained mutatis mutandis through the same arguments as in Lemma \ref{lema1}. The substantial difference stays withing Condition {\em 1}., which needs careful justification. We start by invoking an isometric embedding argument as much as in \cite{anderes2020}, which proves that the metric space $(\mathbb{R}^m, \rho_{1,m})$ is isometrically embeddable into the quasi metric space $({\cal G},d_{\cdot})$, with ${\cal G}$ a Euclidean tree with $\ceil{m/2}$ leaves, and $\ceil{m/2} \ge 3$. Hence, Condition {\em 1}. translates into the fact that the function $d_{\cdot} \mapsto f(d_{\cdot} \; | \btheta)$ is positive definite. The fact that $f$ can by arbitrarily rescaled by $\xi$ as much as in Lemma \ref{lema1} comes from Theorem 2 in \cite{zastavnyi2000positive}.
\end{proof}

\begin{theorem} \label{thm2}
Let ${\cal A}_{\nu}$ be the Askey function defined at (\ref{askey}). Let $m$ be a positive integer, and let ${\cal G}$ be a Euclidean tree with $\ceil{m/2}$ leaves. Let $G_{\alpha,\beta}$ be the mapping defined through Equation (\ref{construction_0}), with 
$\varphi(\cdot \; | \btheta)={\cal A}_{\nu}(\cdot/c_S)$ with $\nu \ge 2m-1$ and $c_S > 0$. Let $\gamma \ge 1$, and call $\alpha= \nu+1$ and $\lambda=\nu+\gamma+1$. 
Consider the special case 
$$ G_{-\alpha,1} (d_{\cdot}(\cdot,\cdot),|\cdot|^2) = \psi(|\cdot|^2 \; |\bvartheta)^{\alpha} {\cal A}_{\lambda} \left ( \frac{d_{\cdot}}{c_S \psi(|\cdot|^2 \; |\bvartheta)}\right ).$$
Then:
\begin{enumerate}
\item if time is linear $(\mathbb{T}=\mathbb{R})$,  $G_{-\alpha,1}$ is positive definite provided $\psi$ is completely monotonic on the positive real line;
\item if time is circular $(\mathbb{T}=\S)$, $G_{-\alpha,1}$ is positive definite provided $\psi$ is the restriction to the interval $[0,\pi]$ of a completely monotone function.
\end{enumerate}
\end{theorem}

\begin{proof}
We consider the function ${\cal A}_{\nu}$ for $\nu \ge 2m-1$. Arguments in \cite{zastavnyi2000positive} show that ${\cal A}_{\nu}(\rho_{1,m}/c_S)$ is positive definite for any positive $c_S$. Hence, we can invoke the isometric embedding argument in the proof of Lemma \ref{lema2} to claim that the mapping $d_{\cdot} \mapsto {\cal A}_{\nu}(d_{\cdot}/c_S)$ is positive definite over a Euclidean tree with $\ceil{m/2}$ leaves. Hence, ${\cal A}_{\nu}$ can be used as the mapping $f$ in Condition {\em 1}. of Lemma \ref{lema2}. As for the mapping $g$ in Condition {\em 2}. of Lemma \ref{lema2}, we let $\widetilde{\bvartheta}=(\nu,\gamma,\bvartheta^\top)^{\top}$. We consider the mapping $g\left (\cdot \; \Big |\xi, \widetilde{\bvartheta}\right )=\xi^{\nu} \{1- \xi / \psi(|\cdot| \; | \widetilde{\bvartheta}) \}_+^{\gamma} $. Given the assumption on the mapping $\psi$, direct inspection proves that $g$ is positive, decreasing and convex on the positive real line, with $\lim_{t \to \infty} g\left (t \; \Big |\xi, \widetilde{\bvartheta}\right )=0$. Hence, P{\'o}lya criterion \citep{polya1949remarks}, $g\left (|\cdot | \; \Big |\xi, \widetilde{\bvartheta}\right )$ satisfies the requirements of Condition {\em 2}. in Lemma \ref{lema2}. Condition {\em 3}. is verified through direct inspection. Hence, we consider $\mu$ in the metric space $([0,\infty), \mathbb{B}, \mu)$ as the Lebesgue measure, and we claim that 
the scale mixture 
\begin{equation} \label{scale}
\int_{0}^{\infty} {\cal A}_{\nu} \left ( \frac{d_{\cdot}}{\xi} \right ) g\left (|\cdot|\; \Big | \xi, \widetilde{\bvartheta}\right ) {\rm d } \xi  
\end{equation}
provides a positive definite function. The proof is completed by invoking the arguments in the proof of Theorem 1 of \cite{Porcu-Bevilacqua-Genton-2}, which shows that the integral above agrees with the mapping $G_{-\alpha,1}$.
\end{proof}

\subsection{The class $G_{\alpha,\beta}$ for $\alpha$ positive and $\beta$ negative}

\begin{theorem} \label{thm3}
Let $m$ be a positive integer. Let ${\cal G}$ be a Euclidean tree with $\ceil{m/2}$ leaves. Let $G_{\alpha,\beta}$ be the mapping defined through Equation (\ref{construction_0}), with $\alpha=m+1$.
Let $\varphi(\cdot \; | \btheta)$ be completely monotonic on the positive real line, and $\psi$ a Bernstein function. Consider the special case 
$$ G_{\alpha,-1} (d_{\cdot}(\cdot,\cdot),|\cdot|^2) = \frac{1} {\psi(|\cdot|^2 \; |\bvartheta)^{\alpha}} \varphi \left ( {d_{\cdot}}{\psi(|\cdot|^2 \; |\bvartheta)}\right ). $$
Then:
\begin{enumerate}
\item if time is linear $(\mathbb{T}=\mathbb{R})$,  $G_{\alpha,-1}$ is positive definite provided $\psi$ is a Bernstein function; 
\item if time is circular $(\mathbb{T}=\S)$, $G_{\alpha,-1}$ is positive definite provided $\psi$ is the restriction to the interval $[0,\pi]$ of a Bernstein function.
\end{enumerate}
\end{theorem}

\begin{proof}
We start by considering Lemma \ref{lema2}, with $f$ as much as in the proof of Theorem \ref{thm2}. As for the function $g$, we now consider $g\left (\cdot \; \Big | \xi, \widetilde{\bvartheta}\right )=\xi^{\nu} \left (1- \xi  \psi \left (|\cdot| \; \Big | \bvartheta \right ) \right )_+^{\gamma} $. Similar arguments as in the proof of Theorem \ref{thm2} prove that Conditions {\em 1}., {\em 2}., and {\em 3}. in Lemma \ref{lema2} are satisfied. Hence, we can consider the scale mixture (\ref{scale}) for which again we can invoke theorem 1 in \cite{Porcu-Bevilacqua-Genton-2} to claim that the mapping 
$$ \widetilde{G}_{\alpha, \nu,\gamma}(d_{\cdot}, |\cdot| \; | \bvartheta  ) = \frac{1}{\psi(\cdot \; | \bvartheta)^{\alpha}} {\cal A}_{\nu+\gamma+1} \Big ( d_{\cdot} \psi(\cdot \; | \bvartheta) \Big )$$ is positive definite. We now consider the sequence of positive definite mappings 
$$  \Bigg \{ \widetilde{G}_{\alpha,\nu+c_n+1}\left ( \frac{d_{\cdot}}{c_n}, |\cdot|  \; \Bigg | \bvartheta \right ) \Bigg \}_{n=1}^{\infty}, $$
where $\{ c_n \}_{n=1}^{\infty}$ is an increasing sequence of real constants with $c_1=1$. Clearly, the sequence converges, for $n$ tending to $\infty$, to the function 
$$ (d_{\cdot},|\cdot|) \mapsto \frac{1}{\psi(\cdot \; | \bvartheta)^{\alpha}} {\rm e}^{-d_{\cdot} \psi(\cdot \; | \bvartheta)}.$$
Hence, the proof is completed by invoking the integral representation (\ref{comp_mon}) for a completely monotone function. 
 \end{proof}

\subsection{Half Spectral Representations}

While spectral representations of positive definite functions on planar surfaces as well as on spheres have been known for quite a long time \citep{schoenberg1942}, for the case of quasi-metric spaces, no analogues of classical results are available. Here, we provide a half spectral characterization for the case where {\em space} is the graph with the Euclidean edges and {\em time} is either the real line or the sphere. 

\begin{theorem} \label{thm-spectral}
Let ${\cal G}$ be a graph with Euclidean edges. Let $D_{{\cal G}}^{d_{\cdot}}$ be the disk of the distance $d_{\cdot}$, that is 
$$D_{{\cal G}}^{d_{\cdot}} = \{d_{\cdot}(\bx,\by) \; : \; \bx,\by \in {\cal G} \}. $$
Let $C: D_{{\cal G}}^{d_{\cdot}} \times [0,\pi] \to \mathbb{R}$ such that $C$ is continuous and bounded. Then, $C(d_{\cdot},d_{G})$ is positive definite if and only if
\begin{equation}
\label{half} C(d_{\cdot,},d_{G}) = \sum_{k=0}^{\infty} \widetilde{C}_{k}(d_{\cdot}) \cos \left ( k d_{G}\right ), \qquad d_{\cdot} \in D_{{\cal G}}^{d_{\cdot}}, \; d_G \in [0,\pi], 
\end{equation}
where the sequence $\{ C_{k}(\cdot) \}_{k=0}^{\infty}$ of continuous functions $C_{k}$ is such that $C_{k}(d_{\cdot})$ is positive definite for all $k=0,1,\ldots$ and additionally $\sum_{k=0}^{\infty} C_{k}(0) < \infty$.
\end{theorem}

\begin{proof}
The sufficient part is a direct consequence of the fact that positive definite functions are a convex cone that is closed under linear combinations. The necessity comes as a direct application of Lemma 3.4 in \cite{berg-porcu} and by noting that the pair $({\cal G},d_{\cdot})$ is a quasi metric space. 
\end{proof}

\section{Tables to Build New Space-Time Covariance Functions} \label{A.4}

\begin{table}[h!]
\caption{{\small Parametric families of fuctions $\varphi$ that can be used in Theorems \ref{thm1} or Theorem \ref{thm2}, and comparison of the parametric ranges for the cases of the quasi-metric spaces $(\mathbb{R}^m, \|\cdot\|)$ and $(\S^m,d_{G})$. If ${\cal G}$ is a Euclidean tree, then the distance $d_{R}$ can be replaced by the geodesic distance, $d_G$.} \label{table1} }
\centering
{\small\linespread{1.2}\selectfont
\begin{tabular}{|cccccc|}
\hline 
&&&&& \\
Family & Expression & $({\cal G},d_{R})$ and  $(\S^m,d_{G})$ & $(\mathbb{R}^m,\|\cdot\|)$ &   Thm \ref{thm1} & Thm \ref{thm2} \\
&&&&& \\
\hline 
&&&&& \\ 
Dagum & $\varphi(r)=1 - \left ( \frac{r^{\beta}}{1+r^{\beta}}\right )^{\tau}  $ & $\beta,\tau \in (0,1]$ & $\beta \in (0,1], \tau \in (0,2]$  & {\sc YES} & {\sc YES} \\
\hline
Gen. Cauchy & $\varphi(r)=(1+r^{\alpha})^{-\beta/\alpha}$ & $\alpha \in (0,1]$, $\beta>0$ & $\alpha \in (0,2]$, $\beta>0$ & {\sc YES} & {\sc YES} \\
\hline 
Schilling & $\varphi(r) = \frac{\left ( 1 - {\rm e}^{-2 \sqrt{r+a}} \right ) }{\sqrt{r+a}}$ & $a>0$ & $a>0$  & {\sc YES} & {\sc YES} \\
\hline
Mat{\'e}rn & Equation (\ref{matern}) & $0 < \nu \le 1/2 $ & $\nu >0$ & {\sc NO} & {\sc YES} \\
\hline
Pow. Exponential & $\varphi(r)=\exp(-r^{\alpha})$ & $\alpha \in (0,1]$ & $\alpha \in (0,2]$ &  {\sc NO} & {\sc YES} \\
\hline
\end{tabular}
}
\end{table}

\begin{table}[h!]
\caption{{\small Parametric families of functions $\psi$ that can be used to implement examples from Gneiting family as in Equation (\ref{construction_0}), Theorem \ref{thm1}, Theorem \ref{thm2}, and the family of Dynamically supported covariance functions as in Theorem \ref{thm3}. \label{table2} }}
\centering
{\small\linespread{1.2}\selectfont
\begin{tabular}{|ccc|}
\hline 
Family & Expression  & Parameter Restrictions \\
&& \\
\hline
Dagum & $\psi(r)=1 + \left ( \frac{r^{\beta}}{1+r^{\beta}}\right )^{\tau}  $ & $\beta,\tau \in (0,1]$  \\
\hline
Gen. Cauchy & $\psi(r)= (1+r^{\alpha})^{\beta/\alpha}$ & $\alpha \in (0,1]$, $\beta \le \alpha $ \\ 
\hline
Power & $\psi(r)=c+ r^{\alpha}$ & $\alpha \in (0,1]$, $c>0$ \\
\hline
\end{tabular}}
\end{table}

\begin{table}[t!]
\caption{{Parametric families of covariance functions that are obtained as application of Theorem \ref{thm-spectral}, part B. Second column reports the analytic expression, where $g$ is any correlation function on ${\cal G, d_{G}}$. An additional condition is required for the fifth example, as detailed through the third column. The fourth column details the differentiability at the origin for the spatial margin. All of the members $C$ in the second column are rescaled so that $C(0,0)=1$.}}\label{tabula_rasa}
\begin{center}
{\scriptsize \begin{tabular}{|c|cc|}
\hline
{\bf Family} & {\bf Analytic expression} & {\bf Parameters range} \\
 & &   \\
\hline
 Negative Binomial & $C(d_{G},d_{R})= \left \{ \frac{1-\varepsilon}{  1- \varepsilon g(d_{R}) \cos d_{G}  } \right \}^{\tau}$ & $\varepsilon \in (0,1)$, $\tau >0$  \\  
\hline
 Multiquadric & $C(d_{G},d_{R})=\frac{(1-\varepsilon)^{2\tau}}{ \left \{ 1+ \varepsilon^2- 2\varepsilon g(d_{R}) \cos d_{G} \right \}^{\tau} }$ & $\varepsilon \in (0,1)$, $\tau >0$  \\  
\hline
 Sine Series & $C(d_{G},d_{R})=  {\rm e}^{ g(d_{R})\cos\theta-1} \left \{ 1 +  g(d_{R}) \cos\theta \right \}/2$   \\  
\hline
Sine Power & $C(d_{G},d_{R})=1-2^{-\alpha} \{1-g(d_{R}) \cos d_{G}\}^{\alpha/2}$ & $\alpha \in (0,2]$  \\
\hline 
 Adapted  & $C(d_{G},d_{R})= \left [ \frac{\{1+g^2(d_{R})\}(1-\varepsilon )}{ 1+g^2(d_{R}) -2 \varepsilon g(d_{R}) \cos d_{G} } \right ]^{\tau}$ & $\varepsilon \in (0,1), \tau >0$  \\  
Multiquadric&&$2g(\cdot)/\{1+g^2(\cdot)\}$ \\
&&corr. function on $\mathbb{R}$  \\
\hline
Poisson & $C(d_{G},d_{R})= \exp \left [\lambda\left \{ \cos d_{G} g(d_{R}) -1  \right \} \right ]$ & $\lambda >0$ \\ 
\hline
\end{tabular} }
\end{center}
\end{table}

 \baselineskip 20 pt

\bibliographystyle{apalike}
\bibliography{bibgeostat}

\begin{thebibliography}{}

\bibitem[***, 2022]{emery-porcu-22blind}
*** (2022).
\newblock Extending the {G}neiting class for modeling spatially isotropic and
  temporally symmetric vector random fields.
\newblock {\em Technical Report. Submitted for publication}.

\bibitem[Alegr{\'\i}a et~al., 2019]{APFM}
Alegr{\'\i}a, A., Porcu, E., Furrer, R., and Mateu, J. (2019).
\newblock Covariance functions for multivariate {G}aussian fields evolving
  temporally over planet earth.
\newblock {\em Stochastic Environmental Research and Risk Assessment},
  33(8-9):1593--1608.

\bibitem[Alsheikh et~al., 2014]{alsheikh2014machine}
Alsheikh, M.~A., Lin, S., Niyato, D., and Tan, H.-P. (2014).
\newblock Machine learning in wireless sensor networks: Algorithms, strategies,
  and applications.
\newblock {\em IEEE Communications Surveys $\&$ Tutorials}, 16(4):1996--2018.

\bibitem[Anderes et~al., 2020]{anderes2020}
Anderes, E., M{\o}ller, J., and Rasmussen, J.~G. (2020).
\newblock Isotropic covariance functions on graphs and their edges.
\newblock {\em Annals of Statistics}, 48(4):2478--2503.

\bibitem[Apanasovich and Genton, 2010]{Apanasovich:Genton:2010}
Apanasovich, T.~V. and Genton, M.~G. (2010).
\newblock Cross-covariance functions for multivariate random fields based on
  latent dimensions.
\newblock {\em Biometrika}, 97:15 --30.

\bibitem[Baddeley et~al., 2017]{baddeley2017stationary}
Baddeley, A., Nair, G., Rakshit, S., and McSwiggan, G. (2017).
\newblock Stationary point processes are uncommon on linear networks.
\newblock {\em Stat}, 6(1):68--78.

\bibitem[Baddeley et~al., 2021]{BADDELEY2021100435}
Baddeley, A., Nair, G., Rakshit, S., McSwiggan, G., and Davies, T.~M. (2021).
\newblock Analysing point patterns on networks -- a review.
\newblock {\em Spatial Statistics}, 42:100435.
\newblock Towards Spatial Data Science.

\bibitem[Berg, 2008]{berg2008stieltjes}
Berg, C. (2008).
\newblock Stieltjes-pick-bernstein-schoenberg and their connection to complete
  monotonicity.
\newblock {\em Positive Definite Functions: From Schoenberg to Space-Time
  Challenges}, pages 15--45.

\bibitem[Berg and Porcu, 2017]{berg-porcu}
Berg, C. and Porcu, E. (2017).
\newblock From {S}choenberg coefficients to {S}choenberg functions.
\newblock {\em Constructive Approximation}, 45:217--241.

\bibitem[Bevilacqua et~al., 2022]{bevilacqua2022unifying}
Bevilacqua, M., Caama{\~n}o-Carrillo, C., and Porcu, E. (2022).
\newblock Unifying compactly supported and {M}at\'ern covariance functions in
  spatial statistics.
\newblock {\em Journal of Multivariate Analysis}, page 104949.

\bibitem[Bevilacqua et~al., 2019]{BFFP}
Bevilacqua, M., Faouzi, T., Furrer, R., and Porcu, E. (2019).
\newblock Estimation and prediction using {G}eneralized {W}endland covariance
  functions under fixed domain asymptotics.
\newblock {\em The Annals of Statistics}, 47(2):828--856.

\bibitem[Chil{\` e}s and Delfiner, 2012]{chiles}
Chil{\` e}s, J. and Delfiner, P. (2012).
\newblock {\em Geostatistics: {M}odeling {S}patial {U}ncertainty}.
\newblock Wiley, New York.

\bibitem[Cressie et~al., 2006]{cressie2006spatial}
Cressie, N., Frey, J., Harch, B., and Smith, M. (2006).
\newblock Spatial prediction on a river network.
\newblock {\em Journal of Agricultural, Biological, and Environmental
  Statistics}, 11(2):127.

\bibitem[de~Valpine et~al., 2017]{de2017programming}
de~Valpine, P., Turek, D., Paciorek, C.~J., Anderson-Bergman, C., Lang, D.~T.,
  and Bodik, R. (2017).
\newblock Programming with models: writing statistical algorithms for general
  model structures with nimble.
\newblock {\em Journal of Computational and Graphical Statistics},
  26(2):403--413.

\bibitem[Deng et~al., 2014]{deng2014ginibre}
Deng, N., Zhou, W., and Haenggi, M. (2014).
\newblock The ginibre point process as a model for wireless networks with
  repulsion.
\newblock {\em IEEE Transactions on Wireless Communications}, 14(1):107--121.

\bibitem[Fonseca and Steel, 2011]{fonseca}
Fonseca, T. C.~O. and Steel, M. F.~J. (2011).
\newblock A general class of nonseparable space-time covariance models.
\newblock {\em Environmetrics}, 22(2):224--242.

\bibitem[Gardner et~al., 2003]{gardner2003predicting}
Gardner, B., Sullivan, P.~J., and Lembo, Jr, A.~J. (2003).
\newblock Predicting stream temperatures: geostatistical model comparison using
  alternative distance metrics.
\newblock {\em Canadian Journal of Fisheries and Aquatic Sciences},
  60(3):344--351.

\bibitem[Gelman et~al., 2014]{gelman2014understanding}
Gelman, A., Hwang, J., and Vehtari, A. (2014).
\newblock Understanding predictive information criteria for bayesian models.
\newblock {\em Statistics and Computing}, 24(6):997--1016.

\bibitem[Genton and Kleiber, 2015]{Genton:Kleiber:2014}
Genton, M.~G. and Kleiber, W. (2015).
\newblock Cross-covariance functions for multivariate geostatistics (with
  discussion).
\newblock {\em Statistical Science}, 30(2):147--163.

\bibitem[Georgopoulos and Hasler, 2014]{georgopoulos2014distributed}
Georgopoulos, L. and Hasler, M. (2014).
\newblock Distributed machine learning in networks by consensus.
\newblock {\em Neurocomputing}, 124:2--12.

\bibitem[Gneiting, 2002a]{Gne:2002b}
Gneiting, T. (2002a).
\newblock Compactly supported correlation functions.
\newblock {\em Jourrnal of Multivariate Analysis}, 83:493--508.

\bibitem[Gneiting, 2002b]{Gneiting:2002}
Gneiting, T. (2002b).
\newblock Stationary covariance functions for space-time data.
\newblock {\em Journal of the American Statistical Association}, 97:590--600.

\bibitem[Hamilton et~al., 2017]{hamilton2017representation}
Hamilton, W.~L., Ying, R., and Leskovec, J. (2017).
\newblock Representation learning on graphs: Methods and applications.
\newblock {\em arXiv preprint arXiv:1709.05584}.

\bibitem[Hauer, 2001]{hauer2001overdispersion}
Hauer, E. (2001).
\newblock Overdispersion in modelling accidents on road sections and in
  empirical bayes estimation.
\newblock {\em Accident Analysis \& Prevention}, 33(6):799--808.

\bibitem[Isaak et~al., 2018]{isaak2018principal}
Isaak, D.~J., Luce, C.~H., Chandler, G.~L., Horan, D.~L., and Wollrab, S.~P.
  (2018).
\newblock Principal components of thermal regimes in mountain river networks.
\newblock {\em Hydrology and Earth System Sciences}, 22(12):6225--6240.

\bibitem[Jones et~al., 1991]{jones1991analysis}
Jones, B., Janssen, L., and Mannering, F. (1991).
\newblock Analysis of the frequency and duration of freeway accidents in
  seattle.
\newblock {\em Accident Analysis \& Prevention}, 23(4):239--255.

\bibitem[Mastrantonio et~al., 2019]{mastrantonio}
Mastrantonio, G., Jona~Lasinio, G., Pollice, A., Capotorti, G., Teodonio, L.,
  Genova, G., and Blasi, C. (2019).
\newblock A hierarchical multivariate spatio-temporal model for clustered
  climate data with annual cycles.
\newblock {\em The Annals of Applied Statistics}, 13(2):797--823.

\bibitem[Menegatto et~al., 2020]{menegatto2020gneiting}
Menegatto, V., Oliveira, C., and Porcu, E. (2020).
\newblock Gneiting class, semi-metric spaces and isometric embeddings.
\newblock {\em Constructive Mathematical Analysis}, 3(2):85--95.

\bibitem[Miaou and Lum, 1993]{miaou1993modeling}
Miaou, S.-P. and Lum, H. (1993).
\newblock Modeling vehicle accidents and highway geometric design
  relationships.
\newblock {\em Accident Analysis \& Prevention}, 25(6):689--709.

\bibitem[Montembeault et~al., 2012]{montembeault2012impact}
Montembeault, M., Joubert, S., Doyon, J., Carrier, J., Gagnon, J.-F., Monchi,
  O., Lungu, O., Belleville, S., and Brambati, S.~M. (2012).
\newblock The impact of aging on gray matter structural covariance networks.
\newblock {\em Neuroimage}, 63(2):754--759.

\bibitem[Moradi and Mateu, 2020]{moradi}
Moradi, M. and Mateu, J. (2020).
\newblock First-and second-order characteristics of spatio-temporal point
  processes on linear networks.
\newblock {\em Journal of Computational and Graphical Statistics},
  29(3):432--443.

\bibitem[Murray et~al., 2010]{murray2010elliptical}
Murray, I., Adams, R., and MacKay, D. (2010).
\newblock Elliptical slice sampling.
\newblock In {\em Proceedings of the Thirteenth International Conference on
  Artificial Intelligence and Statistics}, pages 541--548. JMLR Workshop and
  Conference Proceedings.

\bibitem[Paciorek and Schervish, 2006]{paciorek}
Paciorek, C.~J. and Schervish, M.~J. (2006).
\newblock Spatial modelling using a new class of nonstationary covariance
  functions.
\newblock {\em Environmetrics}, 17(5):483--506.

\bibitem[Peron et~al., 2018]{emery-peron-porcu}
Peron, A., Porcu, E., and Emery, X. (2018).
\newblock Admissible nested covariance models over spheres cross time.
\newblock {\em Stochastic Environmental Research and Risk Assessment},
  32(11):3053--3066.

\bibitem[Perry and Wolfe, 2013]{perry2013point}
Perry, P.~O. and Wolfe, P.~J. (2013).
\newblock Point process modelling for directed interaction networks.
\newblock {\em Journal of the Royal Statistical Society: SERIES B: Statistical
  Methodology}, 75(5):821--849.

\bibitem[Peterson et~al., 2007]{peterson2007geostatistical}
Peterson, E.~E., Theobald, D.~M., and ver Hoef, J.~M. (2007).
\newblock Geostatistical modelling on stream networks: developing valid
  covariance matrices based on hydrologic distance and stream flow.
\newblock {\em Freshwater biology}, 52(2):267--279.

\bibitem[Peterson et~al., 2013]{peterson2013modelling}
Peterson, E.~E., Ver~Hoef, J.~M., Isaak, D.~J., Falke, J.~A., Fortin, M.-J.,
  Jordan, C.~E., McNyset, K., Monestiez, P., Ruesch, A.~S., Sengupta, A.,
  et~al. (2013).
\newblock Modelling dendritic ecological networks in space: an integrated
  network perspective.
\newblock {\em Ecology Letters}, 16(5):707--719.

\bibitem[Pew et~al., 2020]{pew2020justification}
Pew, T., Warr, R.~L., Schultz, G.~G., and Heaton, M. (2020).
\newblock Justification for considering zero-inflated models in crash frequency
  analysis.
\newblock {\em Transportation Research Interdisciplinary Perspectives},
  8:100249.

\bibitem[P{\'o}lya, 1949]{polya1949remarks}
P{\'o}lya, G. (1949).
\newblock Remarks on characteristic functions.
\newblock In {\em Proceedings of the [First] Berkeley Symposium on Mathematical
  Statistics and Probability}, pages 115--123. University of California Press.

\bibitem[Porcu et~al., 2018]{porcu-alegria-furrer}
Porcu, E., Alegr{\'i}a, A., and Furrer, R. (2018).
\newblock Modeling temporally evolving and spatially globally dependent data.
\newblock {\em International Statistical Review}, 86(2):344--377.

\bibitem[Porcu et~al., 2020a]{Porcu-Bevilacqua-Genton-2}
Porcu, E., Bevilacqua, M., and Genton, M.~G. (2020a).
\newblock Space-time covariance functions with dynamical compact supports.
\newblock {\em Statistica Sinica}, 30:719--739.

\bibitem[Porcu et~al., 2020b]{porcu201930}
Porcu, E., Furrer, R., and Nychka, D. (2020b).
\newblock 30 years of space--time covariance functions.
\newblock {\em Wiley Interdisciplinary Reviews: Computational Statistics}.

\bibitem[Porcu et~al., 2006]{Porcu:Gregori:Mateu:2006}
Porcu, E., Gregori, P., and Mateu, J. (2006).
\newblock Nonseparable stationary anisotropic space--time covariance functions.
\newblock {\em Stochastic Environmental Research and Risk Assessment},
  21(2):113--122.

\bibitem[Porcu and Mateu, 2007]{porcu-mateu}
Porcu, E. and Mateu, J. (2007).
\newblock Mixture-based modeling for space-time data.
\newblock {\em Environmetrics}, 18:285--302.

\bibitem[Porcu et~al., 2007]{porcu-mateu-bevilacqua}
Porcu, E., Mateu, J., and Bevilacqua, M. (2007).
\newblock Covariance functions which are stationary or nonstationary in space
  and stationary in time.
\newblock {\em Statistica Neerlandica}, 61(3):358--382.

\bibitem[Porcu et~al., 2010]{porcu-mateu-christakos}
Porcu, E., Mateu, J., and Christakos, G. (2010).
\newblock Quasi-arithmetic means of covariance functions with potential
  applications to space-time data.
\newblock {\em Journal of Multivariate Analysis}, 100(8):1830--1844.

\bibitem[Porcu and Schilling, 2011]{porcu-schilling}
Porcu, E. and Schilling, R. (2011).
\newblock From {S}choenberg to {P}ick-{N}evanlinna: {T}owards a complete
  picture of the variogram class.
\newblock {\em Bernoulli}, 17(1):441--455.

\bibitem[Porcu et~al., 2020c]{porcu2020reduction}
Porcu, E., Senoussi, R., Mendoza, E., and Bevilacqua, M. (2020c).
\newblock Reduction problems and deformation approaches to nonstationary
  covariance functions over spheres.
\newblock {\em Electronic Journal of Statistics}, 14(1):890--916.

\bibitem[Porcu and Zastavnyi, 2011]{Porcu20111293}
Porcu, E. and Zastavnyi, V. (2011).
\newblock Characterization theorems for some classes of covariance functions
  associated to vector valued random fields.
\newblock {\em Journal of Multivariate Analysis}, 102(9):1293--1301.

\bibitem[Porcu and Zastavnyi, 2014]{porcu2014generalized}
Porcu, E. and Zastavnyi, V. (2014).
\newblock Generalized {A}skey functions and their walks through dimensions.
\newblock {\em Expositiones Mathematicae}, 32(2):190--198.

\bibitem[Rakshit et~al., 2017]{BADDELEY}
Rakshit, S., Nair, G., and Baddeley, A. (2017).
\newblock Second-order analysis of point patterns on a network using any
  distance metric.
\newblock {\em Spatial Statistics}, 22:129--154.

\bibitem[Schilling et~al., 2012]{SSV}
Schilling, R., Song, R., and Vondracek, Z. (2012).
\newblock {\em Bernstein {F}unctions. {T}heory and {A}pplications}.
\newblock De Gruyter.

\bibitem[Schlather, 2010]{schlather}
Schlather, M. (2010).
\newblock Some covariance models based on normal scale mixtures.
\newblock {\em Bernoulli}, 16(3):780--797.

\bibitem[Schoenberg, 1938]{schoenberg2}
Schoenberg, I.~J. (1938).
\newblock {M}etric {S}paces and {C}ompletely {M}onotone {F}unctions.
\newblock {\em Annals of Mathematics}, 25(39):811--841.

\bibitem[Schoenberg, 1942]{schoenberg1942}
Schoenberg, I.~J. (1942).
\newblock Positive definite functions on spheres.
\newblock {\em Duke Mathematical Journal}, 9(1):96--108.

\bibitem[Shaby and Wells, 2010]{shaby2010exploring}
Shaby, B. and Wells, M.~T. (2010).
\newblock Exploring an adaptive metropolis algorithm.
\newblock {\em Technical Report}.

\bibitem[Shankar et~al., 1997]{shankar1997modeling}
Shankar, V., Milton, J., and Mannering, F. (1997).
\newblock Modeling accident frequencies as zero-altered probability processes:
  an empirical inquiry.
\newblock {\em Accident Analysis \& Prevention}, 29(6):829--837.

\bibitem[Shirota et~al., 2017]{shirota2017space}
Shirota, S., Gelfand, A.~E., et~al. (2017).
\newblock Space and circular time log gaussian cox processes with application
  to crime event data.
\newblock {\em The Annals of Applied Statistics}, 11(2):481--503.

\bibitem[Stein, 1999]{stein-book}
Stein, M.~L. (1999).
\newblock {\em Statistical {I}nterpolation of {S}patial {D}ata: {S}ome {T}heory
  for {K}riging}.
\newblock Springer, New York.

\bibitem[Stein, 2005]{stein-jasa}
Stein, M.~L. (2005).
\newblock Space-time covariance functions.
\newblock {\em Journal of the American Statistical Association},
  100(469):310--321.

\bibitem[Tang and Zimmerman, 2020]{tang}
Tang, J. and Zimmerman, D. (2020).
\newblock Space-time covariance models on networks with an application on
  streams.
\newblock {\em arXiv:$2009.14745$}.

\bibitem[Ver~Hoef et~al., 2006]{ver2006spatial}
Ver~Hoef, J.~M., Peterson, E., and Theobald, D. (2006).
\newblock Spatial statistical models that use flow and stream distance.
\newblock {\em Environmental and Ecological Statistics}, 13(4):449--464.

\bibitem[Watanabe and Opper, 2010]{watanabe2010asymptotic}
Watanabe, S. and Opper, M. (2010).
\newblock Asymptotic equivalence of {B}ayes cross validation and widely
  applicable information criterion in singular learning theory.
\newblock {\em Journal of Machine Learning Research}, 11(12):3571--3594.

\bibitem[White and Porcu, 2019]{white2019nonseparable}
White, P.~A. and Porcu, E. (2019).
\newblock Nonseparable covariance models on circles cross time: A study of
  mexico city ozone.
\newblock {\em Environmetrics}, 30(5):e2558.

\bibitem[Xiao et~al., 2017]{xiao2017modeling}
Xiao, S., Yan, J., Yang, X., Zha, H., and Chu, S. (2017).
\newblock Modeling the intensity function of point process via recurrent neural
  networks.
\newblock In {\em Proceedings of the AAAI Conference on Artificial
  Intelligence}, volume~31.

\bibitem[Zastavnyi, 2000]{zastavnyi2000positive}
Zastavnyi, V.~P. (2000).
\newblock On positive definiteness of some functions.
\newblock {\em Journal of Multivariate Analysis}, 73(1):55--81.

\bibitem[Zhang, 2004]{zhang2004inconsistent}
Zhang, H. (2004).
\newblock Inconsistent estimation and asymptotically equal interpolations in
  model-based geostatistics.
\newblock {\em Journal of the American Statistical Association},
  99(465):250--261.

\end{thebibliography}

\end{document}